\documentclass[prd,aps,preprint,tightenlines,superscriptaddress]{revtex4}
\usepackage{amsfonts}
\usepackage{amssymb}
\usepackage{mathrsfs}

\usepackage{amsmath}
\usepackage{epsfig}

\begin{document}

\title{General CP Violation in Minimal Left-Right Symmetric
Model\\ and Constraints on the Right-Handed Scale}
\author{Yue Zhang}
\affiliation{Center for High-Energy Physics and Institute of
Theoretical Physics, Peking University, Beijing 100871, China}
 \affiliation{Maryland Center for Fundamental Physics, Department of Physics, University of
Maryland, College Park, Maryland 20742, USA }
\author{Haipeng An}
 \affiliation{Maryland Center for Fundamental Physics, Department of Physics, University of
Maryland, College Park, Maryland 20742, USA }
\author{Xiangdong Ji}
 \affiliation{Maryland Center for Fundamental Physics, Department of Physics, University of
Maryland, College Park, Maryland 20742, USA } \affiliation{Center
for High-Energy Physics and Institute of Theoretical Physics, Peking
University, Beijing 100871, China}
\author{Rabindra N. Mohapatra}
 \affiliation{Maryland Center for Fundamental Physics, Department of Physics, University of
Maryland, College Park, Maryland 20742, USA }
\date{\today}
\begin{abstract}

In minimal left-right symmetric theories, the requirement of parity invariance allows
only one complex phase in the Higgs potential and one in the Yukawa couplings, leading to
a  two-phase theory with both spontaneous and explicit CP violations. We present a
systematic way to solve the right-handed quark mixing matrix analytically in this model
and find that the leading order solution has the same hierarchical structure as the
left-handed CKM matrix with one more CP-violating phase coming from the complex Higgs
vev. Armed with this explicit right-handed mixing matrix, we explore its implications for
flavor changing and conserving processes in detail, low-energy CP-violating observables
in particular. We report an improved lower bound on the $W_R$ mass of 2.5 TeV from
$\Delta M_K$ and $\Delta M_{B}$, and a somewhat higher bound (4 TeV) from kaon decay
parameters $\epsilon$, $\epsilon'$, and neutron electric dipole moment. The new bound on
the flavor-changing neutral Higgs mass is 25 TeV.
\end{abstract}
\maketitle

\section{introduction}

The physics beyond the Standard Model (SM) has been the central focus of high-energy
phenomenology for more than three decades. Many proposals, including supersymmetry,
technicolor, little Higgs, and extra dimensions, have been made and studied thoroughly in
the literature, tests are soon to be made at the Large Hadron Collider (LHC). One of the
earliest proposals, the left-right symmetric model (LRSM), was motivated by the
hypothesis that parity is a perfect symmetry at high-energy, and is broken spontaneously
at low-energy due to the asymmetric vacuum~\cite{lrmodel}. Asymptotic restoration of
parity has a definite aesthetic appeal~\cite{lee}. The model has a number of
 additional attractive features, including a natural explanation of weak
hyper-charge in terms of baryon and lepton numbers, existence of right-handed neutrinos
and entailed seesaw mechanism for neutrino masses, possibility of spontaneous CP
(charge-conjugation-parity) violation, and natural solution for the strong CP problem.
The model can be constrained strongly by low-energy physics and predicts clear signatures
at colliders. It so far remains a decent possibility for new physics.

The LRSM is best constrained at low-energy by flavor-violating mixing and decays,
particularly CP violating observables. In making theoretical predictions, the major
uncertainty comes from the unknown right-handed quark mixing matrix, similar in spirit to
that of the left-handed quark Cabibbo-Kobayashi-Maskawa (CKM) mixing. The new mixing is a
unitary matrix, depending on 9 real parameters: 6 CP violation phases and 3 rotational
angles. All are physical after the left-handed CKM mixing is rotated into a standard
4-parameter form.

Historically, two special CP violation scenarios in LRSM have been considered. The first
one, `` the manifest left-right symmetry", assumes that there is no spontaneous CP
violation i.e. all Higgs vacuum expectation values (vev's) are real. Then the quark mass
matrices are hermitian, and the left- and right-handed quark mixings become identical,
modulo the sign uncertainty of the elements from negative quark masses. The second
scenario, ``pseudomanifest left-right symmetry", assumes that the CP violation comes
entirely from spontaneous symmetry breaking (SSB) of the vacuum and all Yukawa couplings
are real \cite{scpv}. Here the quark mass matrices are complex and symmetric, implying
that the right-handed quark mixing is related to the complex conjugate of the CKM matrix
multiplied by additional CP phases. There are few studies of the model with general CP
violations in the literature \cite{generalcp,kiers}. It has been noted that there are
problems with both manifest and pseudo-manifest scenarios. The manifest LRSM with real
potential and vev's always provides more minimization conditions than the number of
vev's, thus has a fine-tuning problem. In any case, in the non-suspersymmetric LRSM, the
assumption of all parameters of the model being real is not technically natural since
loop corrections lead to one of the parameters being complex. On the other hand, the
pseudo-manifest LRSM, where exact CP is assumed before symmetry breaking and all
couplings in the Higgs potential are real, leads in the decoupling limit to either a
model with light triplet Higgs, which is already excluded by SM precision test, or a two
Higgs doublet model, which is excluded by experiment due to large tree level
flavor-changing neutral current \cite{gorbahn}. Therefore, neither of the two special
scenarios can be realistic.

In a recent paper \cite{ourpaper}, we reported a systematic approach to analytically
solving the right-handed quark mixing in the minimal LRSM where only the requirement of
parity invariance is imposed prior to symmetry breaking, leaving automatically only one
CP phase in the Higgs potential and one in the Yukawa couplings and leading to a theory
with both explicit and spontaneous CP violations. This model therefore falls in-between
the above two extreme cases and is free of the problems described above. Our approach is
based on the observation that in the absence of any fine tuning, $m_t\gg m_b$ implies
that the ratio of the two vev's of the Higgs bi-doublet, $\xi = \kappa'/\kappa$, is small
and is of the order of $m_b/m_t$. In the leading-order in $\xi$, we find a linear
equation for the right-handed quark mixing matrix which can be readily solved. We present
an analytical solution of this equation valid to ${\cal O}(\lambda^3)$, where
$\lambda=\sin\theta_C$ is the Cabibbo mixing parameter. The leading right-handed quark
mixing is nearly the same as the left-handed CKM matrix, except for additional phases
which are fixed by $\xi$, spontaneous CP phase $\alpha$, and the quark masses.

Our work is similar in spirit to the detailed numerical study of the general CP LRSM made
by Kiers et al. \cite{kiers}, with the two vev's ratio $\kappa'/\kappa$ fixed exactly to
$m_b/m_t$. It is interesting that the gross feature of the right-handed CKM was obtained
already in the appendix of that paper, in particular the hierarchical structure of flavor
mixing and the magnitude of the Dirac CP phase. However, realistic studies were made
numerically for lack of an explicit solution with known precision. In fact, much of the
numerical work of Ref.~\cite{kiers} goes into solving the right-handed CKM in the
presence of 11 input parameters, which must be scanned through using Monte Carlo to
obtain the physical quark masses and the left-handed CKM mixing. Because the extensive
nature of numerical study, it is difficult to see some of the physics in a clear way, in
particular, the interplay between the explicit and spontaneous CP violations in physical
observables. Our explicit analytic solution for the right-handed CKM makes extensive
analytical studies simple and straightforward.

In this paper, we first give the detailed method obtaining the analytical solution to
righthanded quark mixing \cite{ourpaper}. With this explicit right-handed mixing, we
study the neutral kaon and B-meson systems and the neutron electric dipole moment (EDM)
to obtain the lower bound on righthanded $W$-boson mass scale. The neutral kaon mass
mixing provides a rigorous lower bound $M_{W_R}>2.5$ TeV, with the use of the new lattice
QCD calculations of the four-quark matrix elements and the strange quark mass. The
indirect CP violation $\epsilon$ receives large contributions from both explicit and
spontaneous CP phases, and from both the gauge-boson box diagram and flavor-changing
neutral Higgs (FCNH). There are strong cancelations among all the contributions, which in
turn constrain severely the relation among the spontaneous CP phase, $M_{W_R}$, and FCNH
mass $M_H$. We use the cancelation condition to fix the spontaneous CP phase, which is
then used to predict the neutron EDM in terms of the model parameters. Using the
experimental bound on the EDM, we obtain a strong lower bound on $M_{W_R}$, which can be
improved with better calculations of the hadronic matrix elements and more precise
experimental data. Furthermore, we study implications of direct CP violation in $K^0$ and
$B^0$ decays. In the former case, a strong lower bound on $M_{W_R}$ is obtained under the
factorization assumption for the four-quark matrix elements. The CP violating observables
in the kaon and B systems and the neutron EDM provide competitive or even stronger bounds
than the well-known kaon mass mixing. We also present a detailed study of the Higgs
sector in the presence of the spontaneous CP phase, including the mass spectrum, the
neutral and charged couplings to the quarks, and the bound on the FCNH mass.

The paper is organized as follows. We first briefly introduce the minimal LRSM in Sec.
II. In Sec. III, we report our method in solving for the righthanded CKM matrix in the
scenario of generic CP violation, including both spontaneous and explicit phases as is
the case with only parity invariance. In Sec. IV, we study the Higgs sector, including
the mass spectrum and Higgs couplings to quarks. In Sec. V, we explore the well-known
neutral kaon mass difference to find out updated constraints on the right-handed $W$ mass
and the FCNH mass. We also consider similar constraints from the neutral B-meson system.
The CP violations in various processes are discussed in Sec. VI, including $\epsilon$,
$\epsilon'$, neutron EDM and CP asymmetry in $B \rightarrow J/\psi K_S$, to constrain the
mass of $W_R$ and the spontaneous CP phase $\alpha$. We find that they place consistently
strong lower bounds on $M_{W_R}$ and $M_H$. We conclude the paper in Sec. VII, with a
table summarizing various constraint on the right-handed scales.

\section{The Minimal Left-Right Symmetric Model}

 The minimal LRSM is based on the gauge group $SU(2)_L\times
SU(2)_R\times U(1)_{B-L}$. Parity is assumed to be a good symmetry at the Lagrangian
density level, and is broken spontaneously by vev's of Higgs fields. The electric charge
formula can be written as a generalized Gell-mann-Nishijima formula including the third
component of the right-handed isospin, $T_{3L}$ and $T_{3R}$, and the difference between
baryon and lepton numbers ~\cite{marshak},
\begin{eqnarray}\label{a1}
Q&=&T_{3L} + T_{3R} + (B-L)/2 \ .
\end{eqnarray}
This in turn gives an explicit explanation for the standard model
$U(1)$-hypercharge in terms of physical quantum numbers rather than an
arbitrarily adjustable quantum number $Y$.

In the matter sector, the left-handed fermions form fundamental representations of the
$SU(2)_L$ gauge group, while right-handed ones form the representations of the $SU(2)_R$
gauge group. The right-handed neutrinos are introduced automatically so that the
right-handed leptons also form doublets under $SU(2)_R$
\begin{eqnarray}
Q_L = \left( \begin{array}{c} u_L \\ d_L\\ \end{array} \right)\in
\left( 2,1,\frac{1}{3}\right)\ ,  \;\;\; Q_R = \left( \begin{array}{c}
u_R \\  d_R  \\
\end{array} \right) \in \left( 1,2,\frac{1}{3} \right)\ ,  \nonumber\\
L_L = \left( \begin{array}{c} \nu_L \\ l_L\\ \end{array} \right)\in
\left(2,1,-1\right)\ ,  \;\;\; L_R = \left( \begin{array}{c} \nu_R \\ l_R\\
\end{array} \right) \in \left( 1,2,-1 \right) \nonumber \ ,
\end{eqnarray}
where the quantum numbers are those of the above gauge groups. From
these, we can
easily write down the gauge-coupled Lagrangian density for fermions,
\begin{eqnarray}\label{gauge}
\mathscr{L}^{\rm fermion}  &=&  {\overline Q}_{Li} \gamma^\mu
\left(i\partial_\mu -
\frac{g_L}{2}\overrightarrow{W}_{L\mu}\cdot
\vec{\tau}-\frac{g'}{6}B_\mu \right)Q_{Li} \nonumber \\
&+& {\overline Q}_{Ri} \gamma^\mu \left(i\partial_\mu -
\frac{g_R}{2}\overrightarrow{W}_{R\mu}\cdot
\vec{\tau}-\frac{g'}{6}B_\mu \right)Q_{Ri}\nonumber\\
&+&{\overline L}_{Li} \gamma^\mu \left(i\partial_\mu -
\frac{g_L}{2}\overrightarrow{W}_{L\mu}\cdot
\vec{\tau}+\frac{g'}{2}B_\mu \right)L_{Li} \nonumber \\
&+& {\overline L}_{Ri} \gamma^\mu \left(i\partial_\mu - \frac{g_R}{2}
\overrightarrow{W}_{R\mu}\cdot \vec{\tau}+\frac{g'}{2}B_\mu \right)L_{Ri} \ , \nonumber
\end{eqnarray}
where the index $i=1,2,3$ labels fermion generation with all the fields being in flavor
eigenstates. $\vec{W}_{R,L\mu}$ and $B_\mu$ are the gauge fields associated with the
above gauge groups, with corresponding couplings $g_L=g_R$ and $g'$. $\vec{\tau}$ are
Pauli matrices for isospins. The right-handed currents couple to the gauge bosons $W_R$
in a way symmetric to the left-handed counterparts.

The Higgs sector contains a bidoublet $\phi$, belonging to the $(2,2,0)$
representation
of the gauge group, which is the left-right symmetric version of the SM
Higgs doublet and two triplets $\Delta_{L,R}$ belonging to
$(3,1,2)$ and $(1,3,2)$,
respectively,
\begin{eqnarray}
\phi &=& \left( \begin{array}{cc} \phi_{1}^0 & \phi_{2}^{+} \\
\phi_{1}^{-} & \phi_{2}^0\\ \end{array} \right), \;\;\; \Delta_L = \left(
\begin{array}{cc}\delta_{L}^+ / \sqrt{2}
& \delta_{L}^{++} \\ \delta_{L}^{0} & -\delta_{L}^+ / \sqrt{2}\\
\end{array} \right), \nonumber \\
\Delta_R &=& \left( \begin{array}{cc} \delta_{R}^+ / \sqrt{2} &
\delta_{R}^{++} \\ \delta_{R}^{0} & -\delta_{R}^+ / \sqrt{2}\\
\end{array} \right) \ .
\end{eqnarray}
The Higgs boson's kinetic energy and coupling to the gauge fields are
canonical,
\begin{eqnarray}\label{f1}
\mathscr{L}^{\rm Higgs~ kin} &=& {\rm Tr}[(D_\mu \Delta_L)^{\dagger}(D^\mu
\Delta_L)]  + {\rm Tr}[(D_\mu \Delta_R)^{\dagger}(D^\mu \Delta_R)] \nonumber\\
&+& {\rm Tr}[(D_\mu \phi)^{\dagger}(D^\mu \phi)] \ ,
\end{eqnarray}
where the covariant derivatives are
\begin{eqnarray}
D_\mu \phi &=& \partial_\mu \phi + i\frac{g_L}{2}\overrightarrow{W}_{L\mu}\cdot
\vec{\tau}\phi - i\frac{g_R}{2} \phi\overrightarrow{W}_{R\mu}\cdot
\vec{\tau} \ , \nonumber\\
D_\mu\Delta_{(L,R)} &=&
\partial_\mu\Delta_{(L,R)} + i\frac{g_{(L,R)}}{2}\left[\overrightarrow{W}_{(L,R)\mu}\cdot
\vec{\tau}, ~\Delta_{(L,R)} \right] + ig'B_\mu \Delta_{(L,R)} \nonumber \ .
\end{eqnarray}
Again, left-right symmetry is explicit.

Like the SM, we use the vev's of neutral Higgs fields to break the gauge
symmetry. The
symmetry group $SU(2)_L\times SU(2)_R\times U(1)_{B-L}$ is first broken
to $SU(2)_L\times
U(1)_{Y}$ by the vev $\langle \Delta_R \rangle = v_R$ at TeV or
multi-TeV scale. Then
electroweak symmetry breaking is induced by the vev's of $\phi$.  With a
generic Higgs
potential, all the neutral components of the Higgs get vev's,
\begin{eqnarray}
\langle \phi \rangle &=& \left( \begin{array}{cc}
\kappa e^{i \alpha_1} & 0 \\ 0 & \kappa' e^{i \alpha_2} \\
\end{array} \right), \nonumber \\
\langle \Delta_L \rangle &=& \left( \begin{array}{cc} 0 & 0 \\ v_L
e^{i \theta_1} & 0\\
\end{array} \right),\ \ \
\langle \Delta_R \rangle = \left(
\begin{array}{cc} 0 & 0 \\ v_R e^{i \theta_2} & 0\\
\end{array} \right) \ ,
\end{eqnarray}
where there are 4 neutral complex components and so generically 4 phases. At first it
appears that 3 of them can be eliminated through gauge symmetry because we have 3
generators, $T_{3L}$, $T_{3R}$, and $B-L$, commuting with the electromagnetic charge
operator $Q$. In reality, however, we can eliminate only 2. If the transformation
parameters associated with the above 3 operators are $\theta_L$, $\theta_R$ and
$\theta_{B-L}$, the Higgs fields transform as,
\begin{eqnarray}
\langle \phi \rangle &\rightarrow&e^{iT_{3L} \theta_L}\langle \phi \rangle e^{-iT_{3R}
\theta_R}\ , \nonumber\\ \langle \Delta_L \rangle
&\rightarrow&e^{iT_{3L} \theta_L}\langle \Delta_L \rangle e^{-iT_{3L} \theta_L} e^{i\theta_{B-L}}
\ , \nonumber\\
\langle \Delta_R \rangle &\rightarrow&e^{iT_{3R} \theta_R}\langle \Delta_R \rangle
e^{-iT_{3R} \theta_R} e^{i\theta_{B-L}} \ .
\end{eqnarray}
This leads a transformation of the vev phases,
\begin{eqnarray}\label{aa2}
\alpha_1&\rightarrow&\alpha_1 + \frac{1}{2}\theta_L - \frac{1}{2}\theta_R \ , \ \
\alpha_2 \rightarrow \alpha_2 -
\frac{1}{2}\theta_L + \frac{1}{2}\theta_R\ ,  \nonumber \\
\theta_{1,2} &\rightarrow&\theta_{1,2} - \theta_{L,R} + \theta_{B-L} \ .
\end{eqnarray}
It is clear that there are two independent combinations of transformation parameters,
allowing removing two phases only. Conventionally the phases of $\kappa$ and $v_R$ are
set to zero, and thus the general form of Higgs vev's is simplified to
\begin{eqnarray}\label{a4}
\langle \phi \rangle &=& \left( \begin{array}{cc}
\kappa & 0 \\ 0 & \kappa' e^{i \alpha} \\
\end{array} \right),\nonumber \\
\langle \Delta_L \rangle &=& \left( \begin{array}{cc} 0 & 0 \\ v_L
e^{i \theta_L} & 0\\
\end{array} \right), \ \ \
\langle \Delta_R \rangle = \left(
\begin{array}{cc} 0 & 0 \\ v_R & 0\\
\end{array} \right) \ .
\end{eqnarray}
If one chooses $v_L=0$, the only remaining phase $\alpha$ is physically relevant.

Since the bidoublet $\phi$ transforms non-trivially under both $SU(2)_L$ and $SU(2)_R$,
the gauge bosons $W_L$ and $W_R$ are not the mass eigenstates after SSB. From (\ref{f1})
we get the $W$-boson mass matrix,
\begin{eqnarray}\label{f2}
  - \mathscr{L}^{\rm W-Mass}= \left(\begin{array}{cc} W^-_{L\mu} & W^-_{R\mu}\\
\end{array}\right) \left(\begin{array}{cc} \frac{1}{2}g^2(\kappa^2+\kappa'^2+2v^2_L) &
-g^2\kappa\kappa'e^{-i\alpha} \\ -g^2\kappa\kappa'e^{i\alpha} & g^2
v^2_R \\ \end{array} \right)\left(\begin{array}{c} W^{+\mu}_L \\
W^{+\mu}_R\\ \end{array}\right) \ ,
\end{eqnarray}
where $W^\pm_{L,R} = (W^1_{L,R}\mp iW^2_{L,R})/\sqrt{2}$, and $g=g_L=g_R$ by parity.
Since $v_L$ is at most on the order of the left-handed neutrino masses and
$\sqrt{\kappa^2+\kappa'^2}$ on the SM scale, $v_L\ll\sqrt{\kappa^2+\kappa'^2}$, and we
can neglect $v^2_L$ in Eq.~(\ref{f2}). The mass matrix can be diagonalized by a unitary
transformation
\begin{equation}\label{W}
\left(\begin{array}{c} W_L^+ \\ W_R^+ \\
\end{array}\right)=\left(\begin{array}{cc} \cos\zeta & -\sin\zeta
e^{i\lambda} \\ \sin\zeta e^{-i\lambda} & \cos\zeta\\
\end{array}\right)\left(\begin{array}{c} W_1^+ \\ W_2^+ \\
\end{array}\right) \ ,
\end{equation}
where $W_1^+$ and $W_2^+$ are mass eigenstates (1 and 2 subscripts shall not be confused
with the Cartesian components of the isospins), with masses $M_{W_1}\simeq M_{W_L} =
g\sqrt{\kappa^2 + \kappa'^2}/\sqrt{2}$ and $M_{W_2}\simeq M_{W_R} = gv_R$. The parameters
$\zeta$ and $\lambda$ are related to the CP phase $\alpha$ and the masses of $W_1$,
$W_2$, $\kappa$ and $\kappa'$
\begin{equation}\label{f3}
\lambda=-\alpha \ , \;\;\;\;\tan\zeta=-\frac{\kappa\kappa'}{v^2_R} \simeq -2 \xi
\left(\frac{M_{W_L}}{M_{W_R}}\right)^2 \ ,
\end{equation}
where again $\xi=\kappa'/\kappa$. If there is no cancelation in generating quark masses,
$\xi$ has a natural size $m_b/m_t$, and thus $\zeta$ is suppressed by both
$(M_{W_L}/M_{W_R})^2$ and $m_b/m_t$ and is smaller than $4\times 10^{-5}$, which is much
smaller than the current experimental bound $~10^{-2}$. Even so this tiny mixing will be
the dominating contribution to the neutron EDM, as we will discuss below. In terms of the
mass eigenstates, the charged current couplings in the quark sector are
\begin{eqnarray}\label{WW}
\mathscr{L}^{\rm W-current} &=& - \frac{g_L}{\sqrt{2}} \bar u_{Li} \gamma^\mu \left(
\cos \zeta W_{1\mu}^+ - \sin \zeta e^{i \lambda} W_{2 \mu}^+ \right) d_{Li} \\
&-& \frac{g_R}{\sqrt{2}} \bar u_{Ri} \gamma^\mu \left( \sin \zeta e^{ - i \lambda} W_{1
\mu}^+ + \cos \zeta W_{2\mu}^+ \right) d_{Ri} + {\rm h.c.} \ .  \nonumber
\end{eqnarray}
The above expression is in quark flavor basis. In the next section, we will consider the
quark mass basis in which it will be modified by the CKM mixing matrices.

\section{Right-handed Quark Mixing Matrix}\label{III}

In this section, we will focus on the quark sector, solving for the right-handed quark
mixing matrix compatible with the observed quark masses and the left-handed mixing. The
solution is valid in the general CP violation scenario in which both explicit and
spontaneous CP breakings are allowed. The only significant assumption we will make is
that bidoublet higgs vev's and related Yukawa couplings have a hierarchical structure and
there is no large cancelation in generating quark masses, and the solution can be made in
a systematic expansion of the relevant small parameter.

The most general Yukawa coupling of the quark fields with the Higgs bidoublet $\phi$ is
given by
\begin{equation}\label{yukawa}
  \mathscr{L}_Y =  \overline{Q}_{Li} (h_{ij} \phi + \tilde h_{ij} \tilde \phi) Q_{Rj} + {\rm h.
  c.}\ ,
\end{equation}
where $\tilde \phi = -i\tau_2 \phi^* i \tau_2$ and flavor indices $i,j=1,2,3$. Parity
symmetry, under which $\phi\leftrightarrow \phi^\dagger$ and $Q_{Li}\leftrightarrow
Q_{Ri}$, constrains $h$ and $\tilde h$ be hermitian matrices. After SSB, the above
Lagrangian density yields the following up- and down-type quark mass matrices,
\begin{eqnarray}
     M_U &=& \kappa h + \kappa' e^{-i\alpha} \tilde h \nonumber \ , \\
     M_D &=& \kappa' e^{i\alpha} h  + \kappa \tilde h\ .
\end{eqnarray}
There are two terms in each, and we {\it assume} that there is no fine-tuned cancelation
to generate a quark mass scale. Since the top quark mass is much larger than that of the
bottom quark, the assumption implies that $h$ and $\tilde{h}$, $\kappa$ and $\kappa'$
should not be on the same order. Without loss of generality, we take $\kappa'\ll\kappa$
and $\tilde{h}\ll h$. To leading order in $\kappa'/\kappa$, we have
\begin{eqnarray}
M_U &\simeq& \kappa h \ ,  \nonumber\\ \label{a6}M_D &=& \kappa' e^{i\alpha} h + \kappa
\tilde{h} \ . \label{mvev}
\end{eqnarray}
The two terms in the down-type quark masses can be on the same order, however. Because of
the flavor independence of the gauge coupling and the hermiticity of $h$, we can work in
the basis where $M_U$ is diagonal,
\begin{eqnarray}
M_U = S_U \left(\begin{array}{ccc} m_u  &  &  \\  & m_c & \\  &  & m_t \\
\end{array}\right)\equiv S_U \widehat M_U \ ,
\label{upquark}
\end{eqnarray}
in which $S_U= {~\rm diag}\{s_u,s_c,s_t\}$ is the sign of the up-type quark masses. It is
present because the eigenvalues of a hermitian matrix can either be positive or negative,
and by convention we take all $m_i$ positive. In this basis, $M_D$ is not diagonal and is
related to $\widehat M_D= {~\rm diag}\{m_d,m_s,m_b\}$ via left- and right-handed CKM
rotations. Since the phase factor $e^{i\alpha}$ in $M_D$ is generically non-zero, $V^{\rm
CKM}_L \neq V^{\rm CKM}_R$.
\begin{equation}
M_D=V^{\rm CKM}_L  \widehat M_D V^{\rm CKM\dagger}_R S_U \ . \label{downquark}
\end{equation}
For simplicity, we will omit the superscript CKM henceforth. From Eqs.~(\ref{mvev}),
(\ref{upquark}) and (\ref{downquark}), we have
\begin{equation}\label{a7}
\kappa \tilde{h}=V_L \widehat M_DV^\dagger_R S_U-\frac{\kappa'}{\kappa}S_U\widehat
M_Ue^{i\alpha} \ .
\end{equation}
Two comments are in order. First, through the phase transformations that are chirally
independent but isospin-dependent, $u_{L,R}^i \rightarrow e^{i\alpha_i} u_{L,R}^i$ and
$d_{L,R}^i \rightarrow e^{i\beta_i} d_{L,R}^i$, one can bring $V_L$ to a standard form
with only 4 parameters (3 rotations and 1 CP violation phase) and the above equation
remains the same. The $\tilde h$ matrix, however, will be subjected to a unitary
transformation and remains hermitian. Second, after the transformation, all parameters in
the unitary matrix $V_R$ must be physical, including 3 rotations and 6 CP-violating
phases.

To make further progress, one uses the hermiticity condition for $\tilde h$, which leads
to the following equation,
\begin{equation}\label{a19}
  \widehat M_D\widehat V_R^\dagger - \widehat V_R \widehat M_D = 2i\xi\sin\alpha ~ V^\dagger_L \widehat M_U S_U
  V_L \ ,
\end{equation}
where $\widehat V_R$ is the quotient between the left and right mixing $V_R = S_U
V_L\widehat V_R$. There are a total of 9 equations in the above expression, which is just
enough to solve 9 parameters in $\widehat V_R$. It is interesting to note that if there
is no spontaneous CP violation, $\alpha=0$, we recover the solution $V_R = S_U V_L S_D$,
in which $S_D= {~\rm diag}\{s_d,s_s,s_b\}$ is the sign matrix for down-type quark masses.
This is just the manifest LRS case.

The above linear equations can be readily solved utilizing the hierarchy between
down-type quark masses. We find an analytical expression for $V_R$ up to order of
$O(\lambda^3)$, where $\lambda$ is sine of Cabibbo angle. We begin directly from
(\ref{a19}) and the left side is anti-hermitian, and we can write it explicitly
\begin{equation}
\left(\begin{array}{ccc} -2i m_d {~\rm Im} \widehat V_{R11} & - m_s \widehat V _{R12} & -
m_b
\widehat V_{R13}\\ m_s\widehat V_{R12}^* & -2i  m_s {~\rm Im}\widehat V_{R22} & - m_b \widehat V_{R23} \\
m_b\widehat V_{R13}^*& m_b\widehat V_{R23}^*&-2i m_b {~\rm Im}
\widehat V_{R33}\\
\end{array}\right) \ ,
\end{equation}
where we have used $m_d\ll m_s\ll m_b$. The right-hand side of Eq. (\ref{a19}) depends on
the physical quark masses, the standard CKM matrix, and the spontaneous CP violation
parameter $\xi \sin\alpha$. Thus we can solve $\widehat V_R$ in terms of these up to
${\cal O}(\lambda^3)$
\begin{eqnarray}
{~\rm Im} \widehat V_{R11}&=&-r\sin\alpha\frac{m_bm_c}{m_dm_t}\lambda^2 \left(s_c+s_t
\frac{m_t}{m_c}A^2\lambda^4\left((1-\rho)^2+\eta^2\right)\right)\ , \\
{~\rm Im} \widehat V_{R22}&=&-r\sin\alpha\frac{m_bm_c}{m_sm_t}\left(s_c+s_t
\frac{m_t}{m_c}A^2\lambda^4\right)\ , \\
{~\rm Im} \widehat V_{R33}&=&-r\sin\alpha s_t\ , \\
\widehat V_{R12}&=&2ir\sin\alpha\frac{m_bm_c}{m_sm_t}\lambda
\left(s_c+s_t\frac{m_t}{m_c}\lambda^4 A^2(1-\rho+i\eta)\right)\ , \\
\widehat V_{R13}&=&-2ir\sin\alpha A\lambda^3(1-\rho+i\eta)s_t\ , \\
\widehat V_{R23}&=&2ir\sin\alpha A\lambda^2 s_t \ ,
\end{eqnarray}
where $r \equiv (m_t/m_b) \xi$, and $\lambda$, $A$, $\rho$, and $\eta$ are Wolfenstein
parameters for $V_L$. The above solution exists only when $|r\sin\alpha|\le 1$, which is
an interesting and unexpected constraint. Since the natural size of $\xi$ is $m_b/m_t$,
$r\sim 1$, allowing angle $\alpha\sim 1$. Given the physical values of various
parameters, we find the following power counting: Im$\widehat V_{R11} \sim \lambda$,
Im$\widehat V_{R22} \sim \lambda$, $\widehat V_{R12} \sim \lambda^2$, $\widehat V_{R13}
\sim \lambda^3$ and $\widehat V_{R23} \sim \lambda^2$. Using unitarity condition for
$\widehat V_R$, we can solve all other elements.

Defining new phases $\sin\theta_i = S_{Dii} {\rm Im} V_{Rii}$, where $i=1,2,3$, we have
up to ${\cal O}(\lambda^3)$,
\begin{eqnarray}
\widehat V_{Rii}&=&S_{Dii} e^{i\theta_i} \ , \nonumber \\
\widehat V_{R21}&=&-s_ds_s\widehat V^*_{R12}e^{i(\theta_1+\theta_2)}\ , \nonumber\\
\widehat V_{R31}&=&-s_ds_b\widehat V^*_{R13}e^{i(\theta_1+\theta_3)}\ , \nonumber\\
\widehat V_{R32}&=&-s_ss_b\widehat V^*_{R23}e^{i(\theta_2+\theta_3)} \ .
\end{eqnarray}
Therefore, we can write the righthanded CKM matrix in a more compact form
\begin{eqnarray}\label{vr}
V_R = P_U \widetilde V_L P_D \ ,
\end{eqnarray}
in which $P_U={\rm diag}(s_u,s_ce^{2i\theta_2},s_te^{2i\theta_3})$, $P_D={\rm
diag}(s_de^{i\theta_1},s_se^{-i\theta_2},s_be^{-i\theta_3})$, and
\begin{equation}\label{vr1}
\widetilde V_L=\left(\begin{array}{ccc} 1-\lambda^2/2 & \lambda &A\lambda^3(\rho-i\eta)\\
-\lambda & 1-\lambda^2/2 &
A\lambda^2e^{-2i\theta_2}\\A\lambda^3(1-\rho-i\eta)&-A\lambda^2e^{2i\theta_2}&1\\
\end{array}\right) \ ,
\end{equation}
which differs from $V_L$ by a small phase in 23 and 32 elements.

The phases $\theta_i$ are functions of parameter $r\sin\alpha$ and the signs of the quark
masses $s_i$ and $\tilde s_i$. Numerically we have,
\begin{eqnarray}\label{theta}
\theta_1&=&-\sin^{-1}[0.31(s_ds_c+0.18s_ds_t)r\sin\alpha]\ , \nonumber\\
\theta_2&=&-\sin^{-1}[0.32(s_ss_c+0.25s_ss_t)r\sin\alpha]\ , \nonumber\\
\theta_3&=&-\sin^{-1}[s_bs_tr\sin\alpha] \ ,
\end{eqnarray}
where we have taken from the Particle Data Group the central values of the quark masses
$m_u=2.7$ MeV, $m_d=5$ MeV, $m_s=98$ MeV, $m_c=1.25$ GeV, $m_b=4.2$ GeV, and $m_t=174$
GeV at scale 2 GeV \cite{PDG}. It shall be noted that since only the quark mass ratios
enter the mixing matrix and the quark masses run multiplicatively, the result is
independent of quark mass scale. The parameters for the left-hand quark mixing are taken
as $\lambda=0.2272$, $A=0.818$, $\rho=0.221$, and $\eta=0.34$.

A few remarks about the above result are in order. First, the hierarchical structure of
the right-handed mixing is similar to that of the left-handed CKM, namely 1-2 mixing is
of order $\lambda$, 1-3 order $\lambda^3$ and 2-3 order $\lambda^2$. Second, every
element now has a substantial CP phase. When $r$ is of order 1, the elements involving
the first two families have CP phases of order $\lambda$, and the phases involving the
third family are of order $1$. These phases are all related to the single spontaneous
CP-violating phase $\alpha$, and generate rich phenomenology for $K$ and $B$ meson
systems as well as the neutron EDM. Finally, from (\ref{vr}) and (\ref{theta}), it is
clear that the final solution is a function of sign bi-products $s_is_j$. We can always
fix one of them, say $s_u$, to be positive, then we are left with $2^5=32$ distinct
sectors. The actual physical choice must be determined by phenomenology, as we will
illustrate in the following sections.

\section{Higgs potential, mass spectrum and couplings}\label{higgs3}

In this section, we discuss several issues related to the Higgs sector. In particular, we
consider the possibility of spontaneous CP violation from the Higgs potential, the mass
spectrum of the Higgs bosons, and the Higgs couplings to the quark sector. The results
are useful for phenomenological studies in the following sections. Some of the results
presented here have appeared in the literature before, and we include them for
completeness.

The most general renormalizable Higgs potential invariant under parity is given
by~\cite{pot}
\begin{eqnarray}
&&\mathscr{V}(\phi, \Delta_L, \Delta_R) = - \mu_1^2 {\rm Tr} (\phi^{\dag} \phi) - \mu_2^2
\left[ {\rm Tr} (\tilde{\phi} \phi^{\dag}) + {\rm Tr} (\tilde{\phi}^{\dag} \phi) \right]
- \mu_3^2 \left[ {\rm Tr} (\Delta_L \Delta_L^{\dag}) + {\rm Tr} (\Delta_R
\Delta_R^{\dag}) \right] \nonumber
\\
&&+ \lambda_1 \left[ {\rm Tr} (\phi^{\dag} \phi) \right]^2 + \lambda_2 \left\{ \left[
{\rm Tr} (\tilde{\phi} \phi^{\dag}) \right]^2 + \left[ {\rm Tr}
(\tilde{\phi}^{\dag} \phi) \right]^2 \right\} \nonumber \\
&&+ \lambda_3 {\rm Tr} (\tilde{\phi} \phi^{\dag}) {\rm Tr} (\tilde{\phi}^{\dag} \phi) +
\lambda_4 {\rm Tr} (\phi^{\dag} \phi) \left[ {\rm Tr} (\tilde{\phi} \phi^{\dag}) + {\rm
Tr}
(\tilde{\phi}^{\dag} \phi) \right]\nonumber \\
&& + \rho_1 \left\{ \left[ {\rm Tr} (\Delta_L \Delta_L^{\dag}) \right]^2 + \left[ {\rm
Tr} (\Delta_R \Delta_R^{\dag}) \right]^2 \right\} \nonumber \\ && + \rho_2 \left[ {\rm
Tr} (\Delta_L \Delta_L) {\rm Tr} (\Delta_L^{\dag} \Delta_L^{\dag}) + {\rm Tr} (\Delta_R
\Delta_R) {\rm Tr} (\Delta_R^{\dag} \Delta_R^{\dag}) \right] \nonumber
\\
&&+ \rho_3 {\rm Tr} (\Delta_L \Delta_L^{\dag}) {\rm Tr} (\Delta_R \Delta_R^{\dag})+
\rho_4 \left[ {\rm Tr} (\Delta_L \Delta_L) {\rm Tr} (\Delta_R^{\dag} \Delta_R^{\dag}) +
{\rm Tr} (\Delta_L^{\dag} \Delta_L^{\dag}) {\rm Tr} (\Delta_R
\Delta_R) \right]  \nonumber \\
&&+ \alpha_1 {\rm Tr} (\phi^{\dag} \phi) \left[ {\rm Tr} (\Delta_L \Delta_L^{\dag}) +
{\rm Tr} (\Delta_R \Delta_R^{\dag})  \right] \nonumber
\\
&&+ \left\{ \alpha_2 e^{i \delta_2} \left[ {\rm Tr} (\tilde{\phi} \phi^{\dag}) {\rm Tr}
(\Delta_L \Delta_L^{\dag}) + {\rm Tr} (\tilde{\phi}^{\dag} \phi) {\rm Tr} (\Delta_R
\Delta_R^{\dag}) \right] + {\rm h.c.}\right\} \nonumber
\\
&&+ \alpha_3 \left[ {\rm Tr}(\phi \phi^{\dag} \Delta_L \Delta_L^{\dag}) + {\rm
Tr}(\phi^{\dag} \phi \Delta_R \Delta_R^{\dag}) \right] + \beta_1 \left[ {\rm Tr}(\phi
\Delta_R \phi^{\dag} \Delta_L^{\dag}) +
{\rm Tr}(\phi^{\dag} \Delta_L \phi \Delta_R^{\dag}) \right] \nonumber \\
&&+ \beta_2 \left[ {\rm Tr}(\tilde{\phi} \Delta_R \phi^{\dag} \Delta_L^{\dag}) + {\rm
Tr}(\tilde{\phi}^{\dag} \Delta_L \phi \Delta_R^{\dag}) \right] + \beta_3 \left[ {\rm
Tr}(\phi \Delta_R \tilde{\phi}^{\dag} \Delta_L^{\dag}) + {\rm Tr}(\phi^{\dag} \Delta_L
\tilde{\phi} \Delta_R^{\dag}) \right] \ ,
\end{eqnarray}
where there are a total of 18 parameters, $\mu^2_{1,2,3}$, $\lambda_{1,2,3,4}$,
$\rho_{1,2,3,4}$, $\alpha_{1,2,3}$, and $\beta_{1,2,3}$.  Due to the left-right symmetry
(LRS), only one of them, $\alpha_2$ can become complex and all other couplings are real.
We have included an explicit phase $e^{i \delta_2}$ in $\alpha_2$, introducing an
explicit CP violation in the Higgs potential.

After SSB, the Higgs fields acquire vev's, and the potential is minimized with respect to
them. The six minimization conditions are
\begin{eqnarray}
\frac{\partial \mathscr{V}}{\partial \kappa} =\frac{\partial \mathscr{V}}{\partial
\kappa'}=\frac{\partial \mathscr{V}}{\partial \alpha}=\frac{\partial
\mathscr{V}}{\partial v_L}= \frac{\partial \mathscr{V}}{\partial v_R}=\frac{\partial
\mathscr{V}}{\partial \theta_L} = 0 \ ,
\end{eqnarray}
which lead to six relations among the vev's and coefficients in the Higgs
potential~\cite{pot, pot1}
\begin{eqnarray}\label{f9}
&&\frac{\mu_1^2}{v_R^2} = \frac{\alpha_1}{2} \left( 1 +
\frac{v_L^2}{v_R^2} \right) - \frac{\alpha_3 \xi^2}{2(1-\xi^2)}
\left( 1 + \frac{v_L^2}{v_R^2} \right) + \left[ \lambda_1 (1+\xi^2)
+
2 \lambda_4 \xi \cos \alpha \right] \epsilon^2 \nonumber \\
&&\;\;\;\;\;\; + \left[ \beta_2 \cos \theta_L - \beta_3 \xi^2
\cos(\theta_L - 2\alpha)
\right] \frac{v_L/v_R}{1-\xi^2},
\end{eqnarray}
\begin{eqnarray}
&&\frac{\mu_2^2}{v_R^2} = \frac{\alpha_2}{2 \cos \alpha} \left[ \cos(
\alpha + \delta_2 )
+ \cos ( \alpha - \delta_2 ) \frac{v_L^2}{v_R^2} \right] + \frac{\alpha_3 \xi}{4
(1-\xi^2)\cos
\alpha} \left( 1 + \frac{v_L^2}{v_R^2} \right) \nonumber \\
&&\;\;\;\;\;\; +\left[ 2 \lambda_2 \xi \cos 2 \alpha + \lambda_3 \xi + \frac{1}{2}
\lambda_4 (1+ \xi^2) \cos \alpha \right]
\frac{\epsilon^2}{\cos \alpha} \nonumber \\
&&\;\;\;\;\;\; + \left[ \beta_1 (1- \xi^2) \cos (\theta_L - \alpha) - 2 \beta_2 \xi
\cos\theta_L + 2 \beta_3 \xi \cos(\theta_L - 2 \alpha) \right] \frac{v_L/v_R}{4(1-
\xi^2)\cos \alpha},
\end{eqnarray}
\begin{eqnarray}
&&\frac{\mu_3^2}{v_R^2} = \rho_1 \left( 1 + \frac{v_L^2}{v_R^2} \right) + \frac{1}{2}
\left[ \alpha_1 (1 + \xi^2) + \alpha_3 \xi^2 \right] \epsilon^2 \nonumber \\
&& + 2 \alpha_2 \left[ \cos( \alpha + \delta_2 ) - \cos ( \alpha - \delta_2 )
\frac{v_L^2}{v_R^2} \right] \frac{\xi \epsilon^2}{1-v_L^2/v_R^2},
\end{eqnarray}
\begin{eqnarray}
&&\left[ (2\rho_1 - \rho_3) - \frac{8 \alpha_2 \xi \epsilon^2 \sin
\alpha \sin \delta_2}{1 - v_L^2/v_R^2} \right]\frac{v_L}{v_R}=
[\beta_1 \xi \cos(\theta_L - \alpha) + \beta_2 \cos \theta_L +
\beta_3 \xi^2 \cos(\theta_L - 2 \alpha)]\epsilon^2 \nonumber\\
\end{eqnarray}
\begin{eqnarray}
&& 0 = \beta_1 \xi \sin(\theta_L - \alpha ) + \beta_2 \sin \theta_L + \beta_3 \xi^2
\sin(\theta_L - 2 \alpha)
\end{eqnarray}
\begin{eqnarray}\label{f14}
&& 2 \alpha_2 (1- \xi^2) (1 - v_L^2/v_R^2) \sin \delta_2 = \left\{ 2
\xi \sin(\theta_L - \alpha) (\beta_2 + \beta_3) \right.   \nonumber
\\ && \left.+ \left[ \sin \theta_L + \xi^2 \sin(\theta_L - 2 \alpha)
\right] \beta_1 \right\} \frac{v_L}{v_R} + \xi \sin \alpha \left[
\alpha_3 (1 + v_L^2/v_R^2) + (4 \lambda_3 - 8 \lambda_2) (1- \xi^2)
\epsilon^2 \right] \nonumber \\
\end{eqnarray}
where $ \epsilon = \kappa/v_R$ represents a hierarchy in symmetry breaking. The above
equations can be solved for the Higgs vev's in terms of the parameters in the Higgs
potential.

Historically, two special cases of the general potential have been studied in the
literature, namely ``manifest" and ``pseudo-manifest" LRS limits. The manifest LRS
assumes real Higgs potential i.e. $\delta_2=0$, and in addition, no spontaneous CP
violation, $\alpha = \theta_L =0$. The only source of CP asymmetry is from the Yukawa
couplings. In this case, the quark mass matrices are hermitian due to parity invariance,
and the left- and right-handed CKM matrices are identical up to quark mass signs. Most
early studies were
made based on this simplification. 
At the level of Higgs potential, this
scenario necessitates fine-tuning: From the neutrino and quark mass
hierarchy, we have
$v_L \ll \kappa' < \kappa \ll v_R$. Taking all the phases to zero in Eqs.
(\ref{f9})-(\ref{f14}), the following relations are found at leading order in
$\epsilon^2$ \cite{pot}
\begin{equation}
\frac{\mu_1^2}{v_R^2} = \frac{\alpha_1}{2} - \frac{\alpha_3
\xi^2}{2(1-\xi^2)}, \ \frac{\mu_2^2}{v_R^2} = \frac{\alpha_2}{2} +
\frac{\alpha_3 \xi^2}{4(1-\xi^2)}, \ \frac{\mu_3^2}{v_R^2} = \rho_1.
\end{equation}
There are three equations for only two vev's $v_R$ and $\xi$, implying a relation among
parameters in the Higgs potential, which can only be achieved through fine-tuning.

On the other hand, pseudo-manifest LRS requires P and CP invariance of the Lagrangian
$(\delta_2=0)$, with the complex vev phase $\alpha$ alone to explain the source of CP
violation in the quark sector. The Higgs potential is real when $\delta_2=0$, but the vev
could be complex. The Yukawa couplings are real and symmetric. The right-handed CKM
matrix is related to the complex conjugate of its left-handed counterpart with additional
diagonal phase matrices multiplied on both sides. However, when Higgs potential is real,
the spontaneous CP phase is proportional to $\sim
\left(\displaystyle\frac{m_{W_L}}{m_{W_R}}\right)^2$ and therefore goes to zero in the
$v_R\to \infty$ limit~\cite{masiero}. If one allowed for fine tuning of parameter, one
can generate a large enough phase~\cite{wolf} but at the price of large flavor changing
neutral current. Phenomenology of these models have been extensively studied in
literature~\cite{pseudo, frere}, and it has been established that this scenario fails to
produce large enough CP asymmetry in $B \rightarrow \psi K$ decay in the decoupling limit
even when the maximum spontaneous CP phase is allowed. Away from the decoupling limit,
the scenario has been ruled out by the sign correlation between $\epsilon$ and the above
B-decay CP asymmetry \cite{pseudo}.

In view of these results, for the minimal LRSMs to be realistic and natural, both
explicit and spontaneous CP phases  must be taken into account. Anyway, as noted this is
precisely what happens in the minimal model. In Ref. \cite{pot1}, an approximate relation
was derived between the spontaneous CP phase $\alpha$ and the explicit CP phase
$\delta_2$ in the Higgs potential,
\begin{equation}
    \alpha \sim
    \sin^{-1}\left(\frac{2|\alpha_2|\sin\delta_2}{\alpha_3 \xi}\right) \ ,
\end{equation}
 where small $\xi$ requires a hierarchy between $\alpha_2$ and
$\alpha_3$, and/or small $\delta_2$. Clearly, when $\delta_2=0$, one has $\alpha\sim 0$.
A pioneering numerical study of the general CP scenario has been made in Ref.
\cite{kiers}. We will consider the Higgs spectrum and coupling in this general case in
the remainder of the section.

\subsection{Higgs Mass Spectrum}

With the Higgs vev's in Eq. (\ref{a4}) and the minimization conditions in Eqs.
(\ref{f9})-(\ref{f14}), the Higgs mass spectrum can be found in the presence of the CP
phase $\delta_2$ in the Higgs potential as well as the spontaneous phase $\alpha$. We
further restrict to the case $\kappa' \ll \kappa$, $v_L=0$. Thus $\theta_L$ becomes
irrelevant and all $\beta_i$ decouple. We will keep only terms linear in $\epsilon$ and
$\xi$ for simplicity.

In the minimal LRSM, there are 20 scalar degrees of freedom in the Higgs fields $\phi$,
$\Delta_L$ and $\Delta_R$, including 2 double-charged, 4 single-charged and 4 complex
neutral Higgs bosons. After SSB, the mass eigenstates are linear combinations of those.
Two single-charged and two real neutral Higgs bosons get absorbed and become longitudinal
components of $W_L$, $W_R$, $Z$ and $Z'$.
\begin{eqnarray}
G_L^+ &=& \left( - \xi e^{-i \alpha} \phi_2^+ + \phi_1^+ \right)\ , \nonumber \\
G_R^+ &=& \left( - \frac{1}{\sqrt{2}} \epsilon \phi_2^+ + \delta_R^+ \right)\ , \nonumber \\
G_{Z'}^0 &=& \sqrt{2} ~{\rm Im}~\delta_R^0 \ , \nonumber \\
G_Z^0 &=& \sqrt{2} ~{\rm Im}~\left( \phi_1^{0*} + \xi e^{-i \alpha}
\phi_2^0 \right) \ ,
\end{eqnarray}
where we have neglected terms of order $\epsilon^2$ and $\xi^2$. Among the remaining 14
fields, only one real and neutral component $h^0$ acquires mass at the electroweak scale
$\kappa$, identified as the SM Higgs boson, while the other Higgs fields have masses of
order $v_R$. The physical Higgs states and their masses are collected in Table
\ref{table}. In the limit $\alpha \rightarrow 0$, $\delta_2 \rightarrow 0$ and $\kappa,
\kappa' \ll v_R$, our results agree with those in Ref. \cite{Duka:1999uc}, except for the
SM Higgs mass.

\begin{table}[hbt]
\begin{tabular}{|c|c|}
\hline
Higgs state & Mass$^2$  \\
\hline
\hline
$h^0 = \sqrt{2} ~{\rm Re}~\left( \phi_1^{0*} + \xi e^{-i \alpha} \phi_2^0
\right)$ & $\left( 4 \lambda_1 - \frac{\alpha_1^2}{\rho_1}
\right) \kappa^2 + \alpha_3 v_R^2 \xi^2$ \\
$H_1^0 = \sqrt{2} ~{\rm Re}~( -\xi e^{i \alpha} \phi_1^{0*} +
\phi_2^0 )$ & $\alpha_3 v_R^2$ \\
$H_2^0 = \sqrt{2} ~{\rm Re}~\delta_R^0$ & $4 \rho_1 v_R^2$ \\
$H_3^0 = \sqrt{2} ~{\rm Re}~\delta_L^0$ & $(\rho_3 - 2 \rho_1) v_R^2$ \\
$A_1^0 = \sqrt{2} ~{\rm Im}~( -\xi e^{i \alpha} \phi_1^{0*} +
\phi_2^0 )$ & $\alpha_3 v_R^2$ \\
$A_2^0 = \sqrt{2} ~{\rm Im}~\delta_L^0$ & $(\rho_3 - 2 \rho_1) v_R^2$ \\
\hline
$H_1^+ = \delta_L^+$ & $(\rho_3 - 2 \rho_1) v_R^2 + \frac{1}{2} \alpha_3 \kappa^2$ \\
$H_2^+ = \phi_2^+ + \xi e^{i \alpha} \phi_1^+ + \frac{1}{\sqrt{2}}\epsilon \delta_R^+$ & $\alpha_3 \left( v_R^2  + \frac{1}{2} \kappa^2 \right)$ \\
\hline
$\delta_R^{++}$ & $4 \rho_2 v_R^2 + \alpha_3 \kappa^2 $ \\
$\delta_L^{++}$ & $( \rho_3 - 2 \rho_1
) v_R^2 + \alpha_3 \kappa^2$ \\
\hline
\end{tabular}
\caption[]{ Physical Higgs states and mass spectrum at leading order in minimal LRSMs. We
assume $v_L =0$ and keep only linear terms in $\epsilon = \kappa/v_R$ and
$\xi=\kappa'/\kappa$. All fields but $h^0$ have masses on the $v_R$ scale. $h^0$ is
identified as the SM Higgs boson.}\label{table} \label{table1}
\end{table}

The calculation of SM Higgs mass is a bit involved and warrants a little further
discussion. From the (tree-level) Higgs potential, one can write down the mass matrix for
8 neutral Higgs components in the basis of $\left\{ {\rm Re}~\phi^0_1,~{\rm
Im}~\phi^0_1,~{\rm Re}~\phi^0_2,~{\rm Im}~\phi^0_2,~{\rm Re}~\delta^0_L,~{\rm
Im}~\delta^0_L,~{\rm Re}~\delta^0_R,~{\rm Im}~\delta^0_R\right\}$. The rows and columns
containing ${\rm Re}~\delta^0_L$, ${\rm Im}~\delta^0_L$ and $G_{Z'} = {\rm
Im}~\delta^0_R$ are already diagonal and hence decouple.

The remaining $5\times5$ sub-matrix $M^2$ is somewhat complicated. Since the major
components have $v_R$-scale masses, we can work in perturbative expansion with respect to
3 parameters: the spontaneous phase $\alpha$, $\xi=\kappa'/\kappa$ and $\epsilon
=\kappa/v_R$. As we shall see later, in order to satisfy the CP constraints and to have
the right-handed scale at TeV, they must be approximately of the same order: $\alpha,
\xi, \epsilon \sim {\cal O}(10^{-2})$. Naively, the SM Higgs mass squared should be of
electroweak scale $\sim \kappa^2 =\epsilon^2 v_R^2$, and we must work up to the order
$\epsilon^2$ in diagonalization: $M^2( \alpha, \xi, \epsilon ) = M_{(0)}^2 + M_{(1)}^2 +
M_{(2)}^2 + \cdots$. At zeroth order, in the basis of $\left\{ {\rm Re}~\phi^0_1,~{\rm
Im}~\phi^0_1,~{\rm Re}~\phi^0_2,~{\rm Im}~\phi^0_2,~{\rm Re}~\delta^0_R\right\}$ \ ,
\begin{equation}
M^2_{(0)}=v_R^2 \left(\begin{array}{ccccc}
0& & & & \\
&0& & & \\
& & \alpha_3& & \\
& & & \alpha_3 & \\
& & & & 4\rho_1 \\
\end{array}\right) \ ,
\end{equation}
which is already diagonal. ${\rm Re}~\phi^0_1$ and ${\rm Im}~\phi^0_1$ have zero mass and
the other three have masses at scale $\sim v_R$. This means ${\rm Re}~\phi^0_1$ and ${\rm
Im}~\phi^0_1$ are the dominant components of the SM Higgs $h^0$ and the Goldstone boson
$G_Z$. At the first order, only the first two rows of $M_{(1)}^2$ are relevant,
\begin{equation}
M^2_{(1)}=v_R^2\left(
\begin{array}{ccccc}
0&0&0&0&2\alpha_1\epsilon \\
0&0&0&\alpha_3\xi&0\\
\end{array}\right) \ ,
\end{equation}
and at the second order, only the upper $2\times2$ block of $M_{(2)}^2$ is needed
\begin{equation}
M^2_{(2)}=v_R^2\left(
\begin{array}{cc}
\alpha_3\xi^2+4\epsilon^2\lambda_1 & 0\\
0 &\alpha_3\xi^2\\
\end{array}\right)\ .
\end{equation}
Using the standard formula in perturbation theory, the SM Higgs boson state and mass up
to the second order in the expansion are
\begin{eqnarray}
h^0 &\simeq& \sqrt{2} ~{\rm Re}\left( \phi_1^{0*} + \xi e^{-i \alpha} \phi_2^0
\right) \ , \nonumber \\
m^2_{h^0}&=&\left(4\lambda_1-\frac{\alpha^2_1}{\rho_1}\right)\kappa^2+\alpha_3\xi^2v_R^2
\ .
\end{eqnarray}
There are two contributions to $m_{h^0}$. One is the usual electroweak breaking mass $4
\lambda_1 \kappa^2$, and the other is the mass shift $\alpha_3 \xi^2 v_R^2 -
\displaystyle\frac{\alpha^2_1}{\rho_1} \kappa^2$, due to additional couplings in the
Higgs potential. Because $\xi = m_b/m_t \sim \epsilon=\kappa/v_R $, the two parts are
roughly comparable when $\lambda_1 \sim \alpha_3 \sim {\cal O}(1)$.

\subsection{Charged and Neutral Currents}\label{coup}

In this subsection, we present the charged and neutral couplings between quarks and Higgs
mass eigenstates. From the Yukawa coupling term (\ref{yukawa}), we can express $h$ and
$\tilde h$ in terms of the vev's and quark mass matrices
\begin{equation}
h = \frac{M_u \kappa - M_d \kappa' e^{- i \alpha}}{\kappa^2 - \kappa^{'2}}, \ \
\tilde h = \frac{M_d \kappa - M_u \kappa' e^{i \alpha}}{\kappa^2 - \kappa^{'2}} \ .
\end{equation}
Using
\begin{equation}
\tilde \phi = \tau_2 \phi^* \tau_2 = \left(
\begin{array}{cc}
\phi_2^{0*} & - \phi_1^+ \\
- \phi_2^- & \phi_1^{0*}
\end{array}\right) \ ,
\end{equation}
one can write
\begin{eqnarray}
\mathscr{L}_Y &=& \overline{Q}_{Li} (h_{ij} \phi + \tilde h_{ij} \tilde \phi) Q_{Rj} +
{\rm h.c.} \equiv \mathscr{ L}_{N} +\mathscr {L}_{C} \ .
\end{eqnarray}
The neutral Higgs-quark coupling part is
\begin{eqnarray}\label{LN}
\mathscr{L}_{N} &=& \bar u_{Li} (h_{ij} \phi_1^0 + \tilde h_{ij} \phi_2^{0*}) u_{Rj} + \bar d_{Li} (h_{ij} \phi_2^0 + \tilde h_{ij} \phi_1^{0*}) d_{Rj} + {\rm h.c.} \nonumber \\
&=& \left( \sqrt{2} G_F \right)^{1/2} \left\{ \bar u_{Li} \Hat M_{Uii} \left[ \left( h^0 - i G_Z^0 \right) - 2 \xi e^{i \alpha} \left( H_1^0 - i A_1^0 \right) \right] u_{Ri} \right. \nonumber \\
& & +\ \left. \bar d_{Li} \widehat M_{Dii} \left[ \left( h^0 + i G_Z^0 \right) - 2 \xi e^{- i \alpha} \left( H_1^0 + i A_1^0 \right) \right] d_{Ri} \right\} \nonumber \\
& & +\ \left( \sqrt{2} G_F \right)^{1/2} \left[ \bar u_{Li} \left(V_L \widehat M_D V_R^\dag\right)_{ij} ( H_1^0 - i A_1^0 ) u_{Rj} \right. \nonumber \\
& & +\ \left. \bar d_{Li} \left(V_L^\dag \Hat M_U V_R\right)_{ij} ( H_1^0 + i A_1^0 ) d_{Rj} \right] + {\rm h.c.}\ ,
\end{eqnarray}
where $V_{L,R}$ are left- and right-handed CKM matrices. $h^0$ has the known SM Higgs
couplings to the quark fields. The second term in Eq. (\ref{LN}) changes the quark
flavors through $H_1^0$ and $A_1^0$ bosons. They are called the flavor changing neutral
Higgs (FCNH) bosons in LRSMs. The dominant contribution comes from the intermediate top,
bottom and charm quark masses. Taking into account the CKM hierarchy, we found that the
transitions from $d$ to $s$ with intermediate charm quark mass and from $b$ to $s$ with
intermediate top quark masse are most significant.

The charged Higgs-quark coupling part of Lagrangian density is
\begin{eqnarray}\label{LC}
\mathscr{L}_{C} &=& \bar u_{Li} (h_{ij} \phi_1^+ - \tilde h_{ij} \phi_2^+) d_{Rj} + \bar d_{Li} (h_{ij} \phi_2^- - \tilde h_{ij} \phi_1^-) u_{Rj} + h.c. \nonumber \\
&=& \left( \sqrt{8} G_F \right)^{1/2} \left[ \bar u_{Li} \left( \Hat M_U V_R - 2 \xi e^{- i \alpha} V_L \widehat M_D \right)_{ij} d_{Rj} H_2^+ \right. \nonumber\\
& & -\ \bar u_{Ri} \left( V_R \widehat M_D - 2 \xi e^{- i \alpha} \Hat M_U V_L \right)_{ij} d_{Lj} H_2^+ \nonumber \\
& & -\ \left. \bar u_{Li} \left( V_L \widehat M_D \right)_{ij} d_{Rj} G_L^+ + \bar u_{Ri}
\left( \Hat M_U V_L \right)_{ij} d_{Lj} G_L^+ \right] + {\rm h.c.} \ .
\end{eqnarray}
Again, the couplings are proportional to quark masses and hence the heavy-quark
contributions stand out. With the above couplings, we will study their contributions to
various flavor changing and conserving processes in the minimal LRSM.

\section{$K^0-\overline{K}^0$ and Neutral B-Meson Mass Mixing}

In this section, we consider the neutral kaon and $B$-meson mass mixing in the minimal
LRSM, using the righthanded quark mixing matrix obtained in the previous section. We
first study the $W_L-W_R$ mixing-box contribution to the $K_L-K_S$ mass difference
$\Delta m_K$ and derive an improved bound (2.5 TeV) on the mass of right-handed gauge
boson $W_R$, using the updated hadronic matrix element and strange quark mass.
Historically, the kaon mass mixing provided the most stringent constraint upper bound
(1.6 TeV) on the mass scale of the right-handed $W_R$ boson \cite{soni}. With our new
right-handed CKM mixing, the conclusion does not change significantly, although in the
literature, quite different mixings have been speculated upon and the result did change
dramatically, and we rule these possibilities out. In the past few years, significant
progress has been made in hadronic physics through lattice QCD simulations, helping to
tighten the bound. We also consider the contribution from the FCNHs and constraint on
their masses. Because the box and FCNH contributions are additive, the bounds are valid
independently. In the last subsection we explore $B_d-\overline B_d$ and $B_s-\overline
B_s$ mass mixing. Because the hadronic contributions arise dominantly from short
distance, the bounds on $M_{W_R}$ and $M_H$ turn out to be significant as well.

It is useful to provide our convention for neutral meson mixing at the beginning. For a
pair of neutral mesons, $|P^0\rangle$ and $|\bar P^0\rangle$, we assume under CP
transformation $CP|P^0\rangle  = |\overline{P}^0\rangle$. If the effective hamiltonian in
the basis of $|P^0\rangle$ and $\overline{P}^0\rangle$ is $H_{ij} = M_{ij} -
i\Gamma_{ij}/2$, a pair of eigenstates are $|P_{1,2}\rangle = p|P^0\rangle \pm q
|\overline{P}^0\rangle$. The ratio $q/p$ is chosen to be $\sqrt{(M^*_{12} -
i\Gamma^*_{12}/2)/(M_{12} - i\Gamma_{12}/2)}$. In the CP symmetric limit, it is possible
to have $q/p=1$, and $P_1$ is then CP-even and $P_2$ is CP-odd. The mass difference is
$M_2 - M_1 = -2{\rm Re}(q/p(M_{12}-i\Gamma_{12}/2))$, and the width difference $\Gamma_2
- \Gamma_1 = 4{\rm Im}(q/p(M_{12}-i\Gamma_{12}/2))$.

\subsection{Kaon Mixing and the Boxing Diagram}

In SM, the leading-order short-distance $\Delta S=2$ process comes from the box diagram
with $W_L$ boson and up-type quark exchanges. Flavor change happens at the vertices via
the CKM mixing matrix. The short distance contribution comes mainly from internal loop
momentum flow at the scales around $c$ and $t$ quark masses, whereas the momentum region
around $W_L$-boson mass is suppressed due to the celebrated Glashow-Iliopoulos-Maiani
(GIM) mechanism. The long-distance contribution comes from one or two up-quark exchanges
and must be calculated using non-perturbative methods. It has been generally accepted
that this latter contribution does not dominate over the short distance one. In fact, a
chiral perturbation calculation \cite{neubert} puts the long distance contribution at
about half of the experimental mass difference.

\begin{figure}[hbt]
\begin{center}
\includegraphics[angle=0, width=0.65\textwidth,height=80pt]{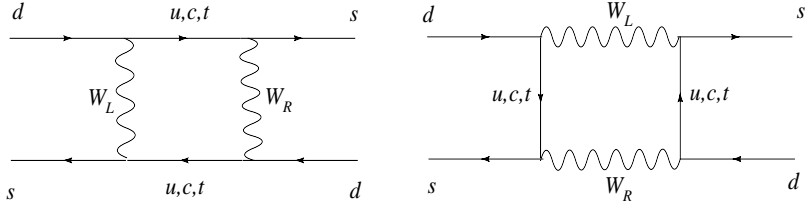}
\end{center}
\caption{Dominant contribution to $K_L$-$K_S$ mass difference in LRS model.} \label{box}
\end{figure}

In the LRSM, there are new box-diagram contributions which turn out to be quite large
\cite{soni, bounds}. The dominant one comes from $W_L-W_R$ interference with one internal
vector-boson being the lefthanded $W_L$ and the other righthanded $W_R$. As such, the
chirality of the internal as well as external quarks must be flipped, and the
contribution is proportional to the internal quark masses, as shown in Fig. \ref{box}.
There is no GIM suppression even in the manifest LRS scenario: $V_L=V_R$. The effective
Lagrangian from the $W_L-W_R$ box-diagram is
\begin{eqnarray}\label{c1}
\mathscr{L}_{LR}&=&-\frac{G_F^2 M_{W_L}^2}{4 \pi^2}2
\eta\sum_{ij}\lambda^{LR}_i\lambda^{RL}_j\sqrt{x_ix_j}\left[(4+\eta
x_ix_j) \right. \nonumber\\
& &\left.  \times I_1(x_i,x_j,\eta) - (1+\eta)I_2(x_i,x_j,\eta)\right]\widehat{O}_{LR}+ {\rm h.c.} \ , \nonumber\\
\end{eqnarray}
in which $\lambda^{\rm LR}_{i}= V^*_{Li2}V_{Ri1}$,
$\eta=\displaystyle{\left(\frac{M_{W_L}}{M_{W_R}}\right)^2}$, and
$x_i=\displaystyle{\left(\frac{m_i}{M_{W_L}}\right)^2}$, $i=u,c,t$. The four quark
operator is
\begin{equation}
  \widehat{O}_{\rm LR}(M_{W_R}^2)=(\overline{s}\mathbb{P}_L d)(\overline{s}\mathbb{P}_Rd) \ ,
\end{equation}
where $\mathbb{P}_{L,R}=(1\mp\gamma_5)/2$ are chiral projection operators. The
loop-related integrals are
\begin{eqnarray}\label{c2}
I_1(x_i,x_j,\eta)&=&\frac{\eta\ln(1/\eta)}{(1-\eta)(1-x_i\eta)(1-x_j\eta)} \nonumber \\
&+& \left[\frac{x_i\ln
x_i}{(x_i-x_j)(1-x_i)(1-x_i\eta)}+(i\leftrightarrow j)\right] \ ,
\nonumber\\
I_2(x_i,x_j,\eta)&=&\frac{\ln(1/\eta)}{(1-\eta)(1-x_i\eta)(1-x_j\eta)}\nonumber \\
&+& \left[\frac{x_i^2\ln
x_i}{(x_i-x_j)(1-x_i)(1-x_i\eta)}+(i\leftrightarrow j)\right] \ .
\nonumber \\
\end{eqnarray}
The above contribution is evaluated in 't Hooft-Feynman gauge. Generally, we have other
contributions from the exchange of single-charged Higgs bosons as well as the Goldstone
bosons if the calculation is not in unitary gauge. In fact, to maintain gauge invariance,
we have also to include the triangle diagrams with the FCNH bosons and the related
self-energy corrections \cite{kaon2}. It turns out that in the gauge we use, the box
diagram with $W_L$ and $W_R$ interchanges alone gives the dominant contribution.

The Wilson coefficient in Eq.~(\ref{c1}) is proportional to the internal quark masses and
the corresponding CKM matrix elements. The up-quark contribution is negligible and the
top-quark contribution is suppressed by the off-diagonal CKM elements. Therefore the
charm-quark exchange dominates the effective interaction which simplifies
\begin{eqnarray}\label{LRLR}
\mathscr{H}_{LR}&=&\frac{G_F^2 M_{W_L}^2}{4 \pi^2} 2 \eta\lambda^{LR}_c\lambda^{RL}_cx_c\left[1+\ln
x_c+\frac{1}{4}\ln\eta\right] \nonumber\\
&&\times \left[(\overline{s}d)^2-(\overline{s}\gamma_5d)^2\right] \ .
\end{eqnarray}
The hadronic matrix element of the four-quark operator is expressed in the vacuum
saturation form,
\begin{eqnarray}
<K_0|\overline{d}(1-\gamma_5)s\overline{d}(1+\gamma_5)s|\overline{K}_0> =-
2M_KF^2_KB_4(\mu)\left(\frac{m_K}{m_s(\mu)+m_d(\mu)}\right)^2 \ ,
\end{eqnarray}
where the kaon decay constant $F_K = 113$~MeV and $B_4=1$ corresponds to vacuum
saturation approximation. In this form, the matrix element diverges in the chiral limit,
an important reason for the enhanced contribution of the box diagram. The correction
factor, $B_4$, can be and has been calculated using lattice QCD. In a recent calculation
~\cite{Bfactor}, the domain-wall fermion was used, and $B_4 = 0.81$ was found at $\mu=2$
GeV in naive dimensional regularization scheme. In the same scheme and scale, the strange
quark mass is $m_s = 98(6)$ MeV, which is smaller than what one has naively expected in
the past.

\subsection{An Improved Lower Bound on $M_{W_R}$}

Considering only the box diagram, the new contribution to the mass difference of
$K_L-K_S$ can be expressed as
\begin{eqnarray}\label{e1}
\Delta m_K &=& 2 ~{\rm Re}~\eta_4(\mu) \langle K^0 | \mathscr{H}^{\Delta S=2}(\mu) |
\bar{K}^0 \rangle\ ,
\end{eqnarray}
where $\eta_4$ is a factor characterizing the QCD radiative correction in scale running
from $M_{W_R}$ to $\mu\sim 2$ GeV~\cite{running, Ecker:1985vv, Buras1}. There are several
enhancement factors here comparing to the SM box diagram. First, due to absence of the
GIM mechanism, the Wilson coefficient is about a factor of 30 larger. Second, the
hadronic matrix element is chirally enhanced by a factor of 20. Finally, the
short-distance QCD correction $\eta_4 =1.4$ gives another enhancement. The only
suppression comes from the difference between the left and right-handed symmetry breaking
scales, $\eta=\left(\displaystyle\frac{M_{W_L}}{M_{W_R}}\right)^2$. Therefore the new
contribution can be approximated by
\begin{equation}
\Delta M_{\rm K-LR}\sim 10^3 \times \left(\frac{M_{W_L}}{M_{W_R}}\right)^2 \times
\Delta M_{\rm K-SM} \ .
\end{equation}
The sign can both be positive or negative depending on the product $s_ts_c$. With the
standard criteria that the new contribution should not exceed the experimental value
\cite{soni}, we find a lower bound for $M_{W_R}$,
\begin{equation}
      M_{W_R} > 2.5~ {\rm TeV} \ .
\end{equation}
On the other hand, the SM contributions from both long and short distances have the same
sign as the experimental number and account for more than one-half of its value.
Therefore, a less conservative bound is obtained if requiring the new physics
contribution is less than one-half of the experimental data. If this new standard is
adopted, the above bound change to 4 TeV. Giving the long-distance uncertainty in $\Delta
M_K$, this bound shall be used at less confidence level. Nonetheless, as we shall see in
the next section, the CP violating observables are providing equally competitive bounds
albeit with a significant hadronic physics uncertainty.

\subsection{Tree-Level FCNH Contribution and A Lower Bound on $M_{H}$}\label{treele}

\begin{figure}[hbt]
\begin{center}
\includegraphics[angle=0, width=10cm]{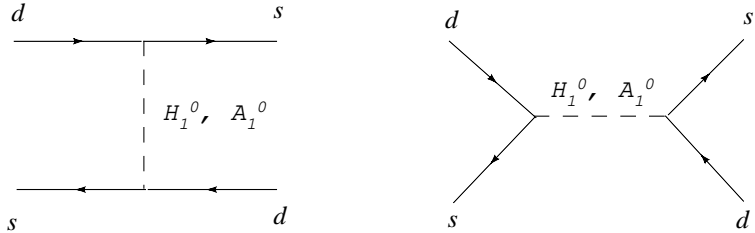}
\end{center}
\caption{$\Delta S=2$ effective interaction induced by flavor-changing neutral Higgses.}
\label{tree}
\end{figure}

In the LRSM, there is also a new contribution to the $K^0-\overline{K}^0$ mixing mediated
by the FCNH. The FCNH boson is a complex field and can be expressed in terms of the two
real fields $H^0_1$ and $A^0_1$. The effective lagrangian follows from Eq.~(\ref{LN})
\begin{eqnarray}\label{27}
&&\mathscr{L}_{FCNH} = \frac{G_F}{\sqrt{2}}
\left[\left(\sum_i\frac{\lambda^{RL}_i+\lambda^{LR}_i}{2}m_i\right)^2
\left[\frac{(\overline{s}d)^2}{m^2_{H^0_1}}-\frac{(\overline{s}\gamma_5d)^2}
{m^2_{A^0_1}}\right] \right. \nonumber\\
&& {\large {\Large }}- \left.\left(\sum_i\frac{\lambda^{RL}_i-\lambda^{LR}_i}{2}
m_i\right)^2\left[\frac{(\overline{s}d)^2}{m^2_{A^0_1}}-
\frac{(\overline{s}\gamma_5d)^2}{m^2_{H^0_1}}\right]\right] \ .
\end{eqnarray}
The corresponding Feynman diagram is shown in Fig. \ref{tree}. According to our previous
discussion, the two scalar fields $H^0_1$ and $A^0_1$ have the same masses, roughly
corresponding to the righthand scale,  $m_{H_1^0}^2 \simeq m_{A_1^0}^2 \simeq \alpha_3
v_R^2$. Therefore, it is convenient to rewrite Eq. (\ref{27}) in a more compact form
\begin{eqnarray}\label{28}
\mathscr{H}_{FCNH} \simeq - \frac{ G_F}{\sqrt{2} m_{H_1^0}^2} \sum_{i,j} m_i m_j
\lambda_i^{LR} \lambda_j^{RL} \left[(\overline{s}d)^2-(\overline{s}\gamma_5d)^2\right] \
.
\end{eqnarray}

It is easy to check that the FCNH and the box diagram contributions have the same sign
because $4 (1 + \ln x_c) + \ln \eta <0$, and thus they cannot cancel each other, even
allowing possible freedom in choosing the quark mass sign.  Therefore, the lower bound on
the righthanded-$W$ boson mass remains. One can also derive a lower bound on the masses
of $H^0_1$ and $A^0_1$ using $\Delta M_K$. A straightforward calculation shows that if
demanding the FCNH contribution is less than the experimental data,
\begin{equation}
    M_{H^0_1}, M_{A^0_1} > 15 ~ {\rm TeV} \ .
\end{equation}
which is about twice as large as in \cite{kiers}. One can obtain this value presumably by
a large $\alpha_3$ parameter in the Higgs potential. However, one cannot make $\alpha_3$
arbitrarily large. As we shall discuss later, large $\alpha_3$ not only causes
naturalness problem, but also leads to a large SM Higgs mass which threatens the
perturbative unitarity \cite{Djouadi:2005gi}.

\subsection{Constraints From Neutral $B_d$ and $B_s$ Mass Mixing}

The physics of $B_d-\overline{B}_d$ and $B_s-\overline{B}_s$ mixing in SM is similar to
that of $K^0-\overline{K}^0$ mixing, coming from the $W$-boson box diagram and FCNH.
However, in the former case, the intermediate top quark contribution dominates almost
entirely due to its mass and CKM couplings. Because of this, the SM calculation can be
done quite accurately with the help of the lattice QCD matrix elements. In fact, in many
global CKM fits, the mass differences $\Delta M_{B}$ and $\Delta M_{Bs}$ have been used
to determine the top CKM couplings $V_{td}$ and $V_{ts}$. However, there are still
appreciable uncertainties in the lattice calculations and global CKM fits, and the beyond
SM physics could contribute as much as $20\%$ of the mass differences without running
into conflict with the present SM calculations and experimental data. We will use this
possible discrepancy as a constraint on the LRSM.


The LRSM contribution to the mixing can be taken directly from (\ref{c1})
\begin{eqnarray}
\mathscr{H}^q_{LR}(\mu_b)&\simeq& \frac{G_F^2 M_{W_L}^2}{4 \pi^2}
2 \eta_S~\eta( V_{tb}^{L*}
V_{tq}^{R} )( V_{tb}^{R*} V_{tq}^{L} ) x_t\left[(4+\eta
x_t^2)  \right. \nonumber\\
& &\left. \times I_1(x_t,x_t,\eta) - (1+\eta)I_2(x_t,x_t,\eta)\right]~\bar b \mathbb{P}_L
q \bar b\mathbb{P}_R q+ {\rm h.c.} \ ,
\end{eqnarray}
where the leading-logarithmic running factor $\eta_S=2.112$ \cite{running} and the
functions $I_i$ are defined in Eq.(\ref{c2}). The hadronic matrix element can be defined
using a factorization,
\begin{eqnarray}
<B_q|\overline{q} \mathbb{P}_L b \overline{q} \mathbb{P}_R b |\overline{B}_q> = - M_{B_q}
f^2_{B_q} B^q_4(\mu) \left[ \frac{1}{12} + \frac{1}{2} \left( \frac{m_{B_q}}{m_b + m_q}
\right)^2 \right] \ ,
\end{eqnarray}
where the minus arises from our definition of the CP transformation for the meson states.
The decay constants have been calculated in lattice QCD: $f_{B_d} = 216 $~MeV, $f_{B_s}=
1.20f_{B_s}$~\cite{lattice} and the non-perturbative $B$-factors are $B_4^d = 1.16$ and
$B_4^s = 1.17$~\cite{lattice2}. The ratio of the new contribution to the SM one is  $\sim
10^2 M_{W_L}^2/M_{W_R}^2$, which is smaller than the kaon mixing case in the absence of
chiral enhancement.

\begin{figure}[hbt]
\begin{center}
\includegraphics[angle=0, width=8cm]{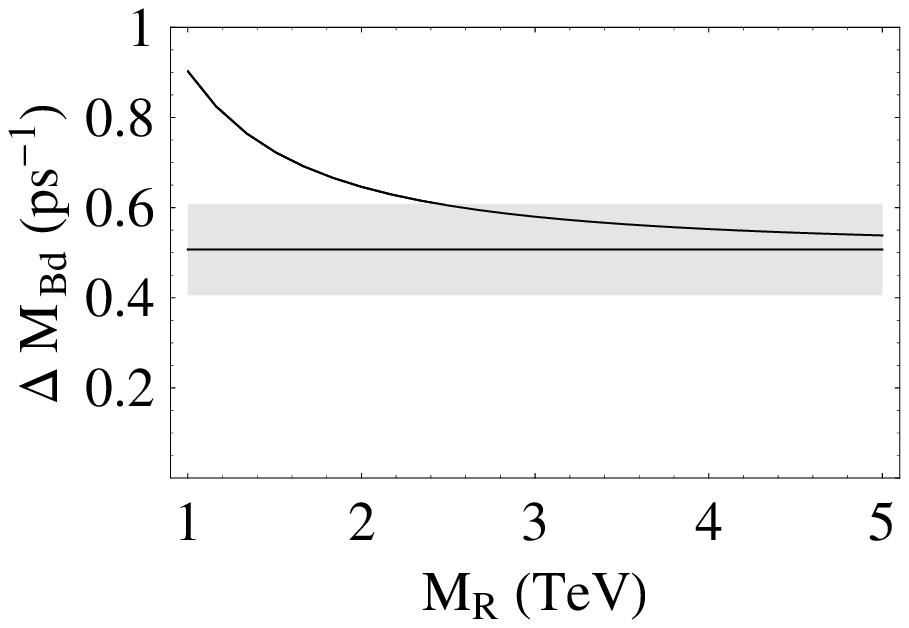}
\includegraphics[angle=0, width=8cm]{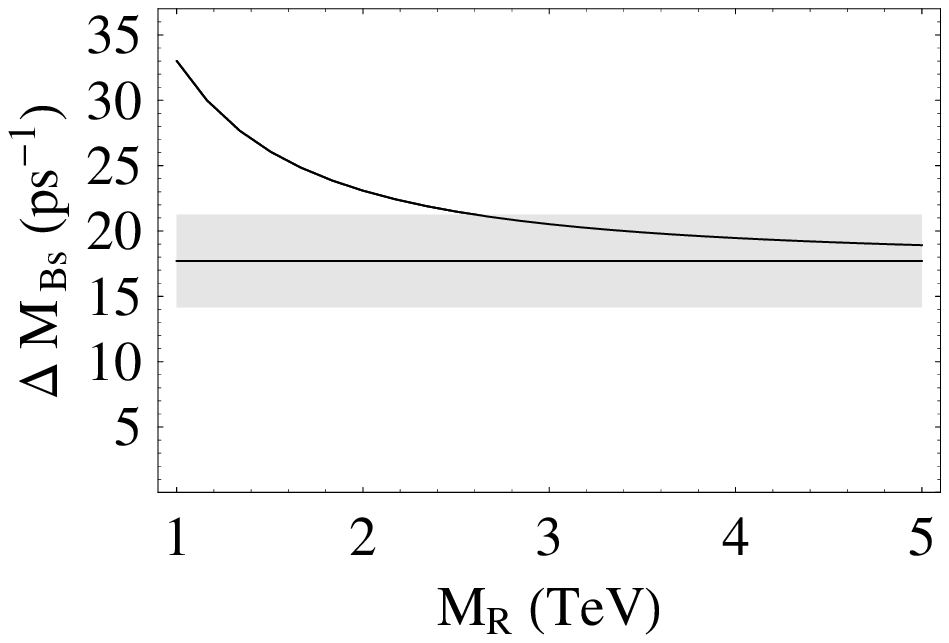}
\end{center}
\caption{$B_d$-$\overline B_d$ and $B_s$-$\overline B_s$ mass differences from the
$W_L-W_R$ box diagram with top-quark intermediate state plus the SM contribution. The
shaded regions are within $20\%$ of the experimental values.} \label{BBB}
\end{figure}

The experimental values for the mass differences of $B_d$ and $B_s$ are
$(0.507\pm0.12)$ps$^{-1}$ and $(17.77\pm0.12)$ps$^{-1}$, respectively~\cite{PDG}, with
central values shown as horizontal lines in Fig. \ref{BBB}. The shades around the central
values are within $20\%$, and are considered as the combined experimental and SM theory
error. In the same figure, we also plot the mass differences as a function of $M_{W_R}$,
calculated as a sum of the LRSM contribution and the experimental central values. The
agreement between experiment and theory is reached only when $M_{W_R}$ is larger than 2.5
TeV. The sign of the LRSM contribution is related to that of $s_ds_b$ and $s_ss_b$, which
we have chosen to be +1. If taking as $-1$, approaching to the central value comes from
below.

Thus the constraint on the $M_{W_R}$ mass from the neutral $B$-meson mixing is roughly
comparable to the kaon case, due to a better theoretical understanding of the SM physics.
The future improvement can come from a better determination of $V_{td}$ from other
sources and a better determination of $B$-parameter and decay constants.

We have also studied the constraint on the FCNH mass, $M_H$. We find that from $\Delta
M_{B_d}$, the bound is 12 TeV, and for $\Delta M_{B_s}$, the bound is 25 TeV.

\section{CP-violating Observables in LRSM}

As mentioned before, in a generic yet minimal LRSM we have both explicit and spontaneous
CP violations. A prominent feature of right-handed quark mixing $V_R$ obtained in the
previous section is its phases, which are entirely determined by the Dirac phase
$\delta_{\rm CP}$ and the spontaneous CP phase $\alpha$. These physical phases generate
interesting effects in various CP-violating observables. The phenomenology of CP
violation in LRSM is rich, which in turn constraints the model severely. In this section,
we will explore a number of CP violating observables including $\epsilon$, $\epsilon'$,
neutron EDM and CP asymmetry in $B \rightarrow J/\psi K_S$, to place constrains on the
mass of $W_R$ as well as the spontaneous CP phase $\alpha$. Some results here have
appeared before in a Rapid Communication paper \cite{ourpaper} and there are important
updates in this more extensive study.

The indirect CP violation $\epsilon$ receives large contributions from both explicit and
spontaneous CP phases. Unless there is a strong cancelation, the right-handed $W_R$ mass
must be larger than 15 TeV. We use the cancelation condition to fix the spontaneous CP
phase, which is then used to predict the neutron EDM. Using the experimental bound on the
EDM, we obtain a strong lower bound on $M_{W_R}$, which can be improved with better
calculations of the hadronic matrix elements and more precise experimental data. We
obtain a strong lower bound on $M_{W_R}$ from the direct CP violation parameter
$\epsilon'$, calculated under the factorization assumption for the four-quark matrix
elements. Therefore, we find that the CP violating observables in the kaon system and
neutron EDM provide competitive bounds as the well-known kaon mass mixing. These bounds
can be improved further with better knowledge of the non-perturbative hadronic physics.

\subsection{Indirect CP violation $\epsilon$ in Kaon Decay}

We first study the CP violating parameter $\epsilon$ in kaon mixng. This indirect CP
violation parameter is related to the flavor mixing interaction by,
\begin{eqnarray}\label{ep}
\epsilon =  -\frac{e^{i \pi/4}}{\sqrt{2}} \frac{~{\rm Im} \langle K^0 |
\mathscr{H}^{\Delta S=2} | \overline{K}^0 \rangle}{ \Delta m_K } \ ,
\end{eqnarray}
where we have neglected the direct CP contribution $\xi_0$ from kaon decay, which can be
justified posteriori \cite{kiers}. In LRSM, according to the previous section, both the
$W_L-W_R$ box diagram in Eq.~(\ref{LRLR}) and the tree level FCNH exchange in
Eq.~(\ref{28}) can make significant contributions. In the present case, their signs can
be different due to both charm and top quark contributions, in contrast to the mass
mixing \cite{kiers}. For simplicity, we ignore that latter contribution and consider the
constraint from the box diagram alone.

There are two sources of CP phases in $V_R$ which enter the $W_L-W_R$ box diagram: the
Dirac phase $\delta_{\rm CP}$ inherited from $V_L$, and the spontaneous phase $\alpha$.
In the manifest LRS case, only $\delta_{\rm CP}$ is present and there is a very tight
lower bound on mass of $W_R$ which we find no lighter than $15-20$ TeV (see below). If
the spontaneous CP phase is also present, one can seek for certain cancelation between
the two contributions to lower the bound on $M_{W_R}$. In fact, one can roughly estimate
the size of $r \sin \alpha$ for a cancelation. The Dirac phase $\delta_{\rm CP}$ appears
in the expression $\epsilon$ proportional to $V^{R*}_{ts} V^R_{td} \sim \lambda^5 \sin
\delta_{CP}$ via top quark exchange in the box diagram, while the spontaneous CP phase
contributes through $V^{R*}_{cs} V^R_{cd} \sim \lambda^2 r \sin \alpha$. Hence $r \sin
\alpha$ should be of order $\lambda^3 \sin \delta_{CP} \sim 0.01$ when cancelation
happens.

\begin{figure}[hbt]
\begin{center}
\includegraphics[angle=0, width=0.35\textwidth,height=96pt]{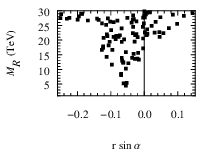}
\includegraphics[angle=0, width=0.35\textwidth,height=96pt]{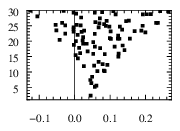}
\includegraphics[angle=0, width=0.35\textwidth,height=96pt]{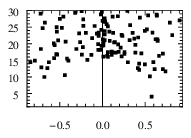}
\includegraphics[angle=0, width=0.35\textwidth,height=96pt]{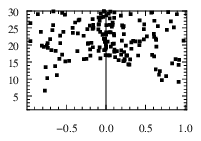}
\end{center}
\caption{Typical scenarios from the $\epsilon$ constraint: The first and two figures
correspond to $s_d = s_s = 1$ and $s_d = s_s = -1$, respectively (small $r \sin \alpha$
solution). The third and fourth correspond to $s_d = - s_s = 1$ and $s_d = - s_s = - 1$
respectively (large $r \sin \alpha$ solution). In all cases, $s_u =  s_c =  s_t =  s_b =
1$. } \label{e}
\end{figure}

The present experiment value is $|\epsilon|_{{\rm expt}} = (2.232 \pm 0.007)\times
10^{-3}$ \cite{PDG}. In SM, $\epsilon$ can be calculated quite accurately because the top
quark dominates the box diagram and the main contribution is due to short-distance QCD
physics. The only large uncertainty comes from the CKM matrix element $V_{td}$ and the
hadronic matrix element related to $K^0-\overline{K}^0$ mixing. Because of this, we
assume that the new contribution accounts for less than 1/4 of the experimental value.

Using the box-diagram result in the previous section, we find an approximate expression
for $\epsilon_{LR}$ valid for $M_{W_R}>200$ GeV \ ,
\begin{equation}\label{CP1}
 \epsilon_{LR} = 1.58 \left( \frac{1~{\rm TeV}}{M_{W_R}} \right)^2 s_s s_d ~{\rm Im}
 \left[ g(M_{W_R}, \theta_2, \theta_3) e^{-i(\theta_1 + \theta_2)} \right]  \ ,
\end{equation}
where
\begin{eqnarray}\label{cp}
g(M_{W_R}, \theta_2, \theta_3) &=& -2.22 + \left[ 0.076 +
(0.030 + 0.013 i)~s_c s_t \right. \nonumber \\
& &\left. \times \cos 2 (\theta_2 - \theta_3) \right] \ln \left( \frac{M_{W_L}}{M_{W_R}
}
\right)^2 \ ,
\end{eqnarray}
where $\theta_i$ are given in Eq.~(\ref{theta}) and the QCD running correction has been
taken into account.  When $\theta_i=0$, the dominant CP violating contribution comes from
charm-top interference in the box diagram.

We search for the allowed region in $M_{W_R}$-$r\sin\alpha$ plane shown in Fig. \ref{e}.
The spontaneous CP violation effect can always cancel the effect of the Dirac phase, thus
$\epsilon$ itself places no bound on $M_{W_R}$. The cancelation depends on choices of the
quark mass signs. The 32 choices of the signs (for fixed $s_u=1$) roughly correspond to
two scenarios: small $r \sin \alpha$ solutions for $s_s=s_d$ and large $r \sin \alpha$
solutions for $s_s = - s_d$. The existence of two scenarios can be readily seen through
the above approximate expression. According to (\ref{theta}), $\theta_1$ and $\theta_2$
are similar in size and their relative sign depends on the bi-product $s_d s_s$. When
$s_s=s_d$, $\theta_1$ and $\theta_2$ have the same sign and add constructively in the
exponential in (\ref{CP1}). So $\alpha$ itself must be small in order for $\theta_1 +
\theta_2$ to cancel the phase in $g(M_{W_R}, \theta_2, \theta_3)$. This generates the
small $r \sin \alpha$ solutions. On the other hand, if $s_s=-s_d$, $\theta_1+\theta_2$
cancels and is proportional to $r \sin \alpha$ multiplied with a small coefficient. So $r
\sin\alpha$ has to be large to cancel the phase in $g(M_{W_R}, \theta_2, \theta_3)$. This
corresponds to the large $r \sin \alpha$ solutions. For the small $r \sin \alpha$ case,
we find $r \sin\alpha \simeq \pm 0.05$, but for large $r \sin \alpha$, it can take
several different positive and negative values. As we will see in the next subsection, if
one includes the constraint from neutron EDM, only small $r\sin\alpha$ is
phenomenologically viable.

\begin{figure}[hbt]
\begin{center}
\includegraphics[angle=0, width=0.50\textwidth]{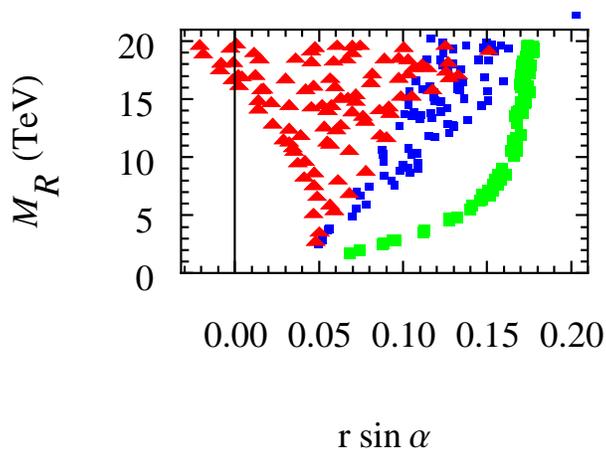}
\end{center}
\caption{Constraint on $M_{W_R}$ and the spontaneous phase $\alpha$ from kaon decay
parameter $\epsilon$, with $s_d=s_s=-1$ and all other $s_i =1$. Red triangles are for
$M_H=\infty$, blue squares are for $M_H=75$ TeV, and green dots for $M_H=20$ TeV.}
\label{fig:fcnh}
\end{figure}

Of course, one has to consider the FCNH contribution which has been known to be large
\cite{Pospelov:1996fq}. In fact, with just the Dirac CP phase in the FCNH contribution,
$\epsilon$ places a limit on the Higgs mass on the order of 100 TeV. With the new
spontaneous CP contribution, there is a possibility of cancelation. In fact, one can make
similar plots as in Fig. \ref{e}, in which the $\epsilon$ bound can be satisfied even for
very low $M_H \sim 1$ TeV. However, the required $r\sin\alpha$ for the cancelation, $ \pm
0.2$, is very different from that needed for the box diagram. The conclusion is that
there is no bound on $M_H$ coming from $\epsilon$ when FCNH contribution is considered
alone.

Because of the conflict in the spontaneous CP phase required for individual cancelations
in the box and FCNH contributions, one might expects their combined contribution places a
joint bound on $M_{W_R}$ and $M_H$. This, however, is not the case, because the two
contributions again cancel each other, as was first found in \cite{kiers}. In fact, with
the analytical solution for the right-handed mixing, we arrive at an even stronger
conclusion: For any given pair of $M_{W_R}$ and $M_H$, we can always find a $r\sin\alpha$
such that the total contribution to $\epsilon$ vanishes. This situation is extremely
interesting, because it implies that $\epsilon$ itself, unlike the kaon mass difference,
is completely useless in constraining the individual parameters in the new contribution.
It does, however, provide a correlation among different parameters, as shown in Fig.
\ref{fig:fcnh}, where for several different values of $M_H=\infty, 75, 25$ TeV, we
plotted the allowed regions in the $M_R$ and $r\sin\alpha$ plane. Because of the FCNH
contribution, the pattern of the correlation changes considerably as $M_H$ changes. The
general trend is that the spontaneous CP parameter $r\sin\alpha$ increases toward 0.2 as
$M_H$ becomes smaller, consistent with the cancelation pattern in FCNH contribution to
$\epsilon$ found above.

\subsection{Neutron EDM}

The neutron EDM imposes another constraint on the new CP phases in $V_R$ and $M_{W_R}$.
Non-zero EDM implies both P and T (or equivalently CP in local quantum field theories)
violations. At the quark level, sources of flavor-neutral CP violation are mainly from
the penguin diagrams in the SM, and from the tree level $W_L-W_R$ exchange in the LRSM
\cite{edm, hehe}. Generally, there are several contributions to the neutron EDM,
including valence quark EDM, quark chromomagnetic dipole moment (CDM) induced EDM,
dimension-6 pure gluonic operator contribution, as well as the contributions at hadronic
level. The present experimental upper bound on neutron EDM is $3.1\times 10^{-26} e$ cm
\cite{PDG}.

In the SM, contributions to the neutron EDM mainly come from the CP violating penguin
diagrams. The flavor changing nature of CKM CP violation means that the leading
contribution is at least second order in weak interaction ($\sim G_F^2$). The predicted
neutron EDM is well within the experimental bound---about $10^{-33} e$ cm.

\begin{figure}[hbt]
\begin{center}
\includegraphics[width=5cm]{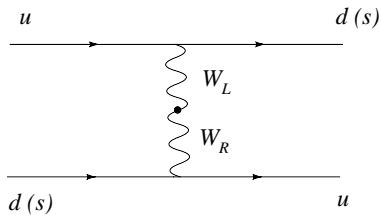}
\end{center}
\caption{Dominant quark-level effective operators contributing to neutron EDM in LRSM.}
\label{edm1}
\end{figure}

In the LRSM, the flavor-conserving CP violating four-quark operator arises from the
tree-level diagram with $W_L-W_R$ mixing exchange, as shown in Fig. \ref{edm1}
\begin{eqnarray}\label{edma}
\mathscr{L}_{uq\rightarrow uq}
&=& - 2 \sqrt{2} G_F \sin \zeta e^{- i \alpha} V^L_{uq} V_{uq}^{R*} \left( O_-^q - O_+^q \right) + {\rm h.c.} \ ,
\end{eqnarray}
where $q=d,s$, and
\begin{eqnarray}\label{edma}
O^q_+ &=& \bar u \gamma^\mu \mathbb{P}_L q \bar q \gamma_\mu \mathbb{P}_R u - \frac{2}{3} \bar u \mathbb{P}_R u \bar q \mathbb{P}_L q \ , \nonumber \\
O^q_- &=& \frac{2}{3} \bar u \mathbb{P}_R u \bar q \mathbb{P}_L q \ .
\end{eqnarray}
At low energy, short-distance QCD effect enhances the operator $O^q_-$ and suppresses
$O^q_+$. The effective Lagrangian reduces to \cite{he}
\begin{eqnarray}
\mathscr{L}_{ud(s)\rightarrow ud(s)} &=& - i \frac{2 \sqrt{2} G_F}{3} \eta_- \sin \zeta
\left[ {\rm Im}\left( e^{- i \alpha} V^L_{ud} V_{ud}^{R*} \right) \left(\bar u \gamma_5 u
\bar d d - \bar u u \bar d \gamma_5 d \right) \right. \nonumber \\ && +\ \left. {\rm
Im}\left( e^{- i \alpha} V^L_{us} V_{us}^{R*} \right) \left(\bar u \gamma_5 u \bar s s
-\bar u  u \bar s \gamma_5s \right) \right] \ ,
\end{eqnarray}
where the leading-log QCD factor $\eta_- = \displaystyle{\left( \frac{\alpha_S(\mu^2)}
{\alpha_S(M_{W_L}^2)} \right)^{\frac{8}{9}}} \simeq 3.5$. The CP violating $\pi n n$
coupling is proportional to the hadronic matrix element $ \bar g_{\pi nn} = \langle \pi n
| \mathscr{L}_{ud(s)\rightarrow ud(s)} | n \rangle $ which, in the factorization
approximation, is
\begin{eqnarray} \langle \pi n | \bar
q \gamma_5 q~\bar q' q' | n \rangle &\simeq& \langle \pi | \bar q \gamma_5 q | 0 \rangle
\langle n |\bar q' q' | n \rangle \ ,
\end{eqnarray}
for $q=u,d$ and $q'=u,d,s$. From the SSB of the chiral symmetry, $2 m_u \langle \pi |
\bar u \gamma_5 u | 0 \rangle = - 2 m_d \langle \pi |\bar d \gamma_5 d | 0 \rangle = - i
F_\pi m_\pi^2$, with $F_\pi=93$ MeV. $\langle n |\bar u u | n \rangle \simeq 4$ and $
\langle n |\bar d d | n \rangle \simeq 5$ can be fixed from the neutron-proton mass
difference and the $\pi N$ $\sigma$-term: $\sigma_N = \displaystyle \frac{1}{2} (m_u +
m_d) \langle n |\bar u u + \bar d d | n \rangle \simeq 45~{\rm MeV}$ \cite{dynamics}, where we
use quark masses $m_u=2.7$ MeV, $m_d=5.0$ MeV at $\mu=2$ GeV. We neglect $\langle n
|\bar s s | n \rangle \ll \langle n |\bar uu | n \rangle $.

\begin{figure}[hbt]
\begin{center}
\includegraphics[width=8cm, height=4.5cm]{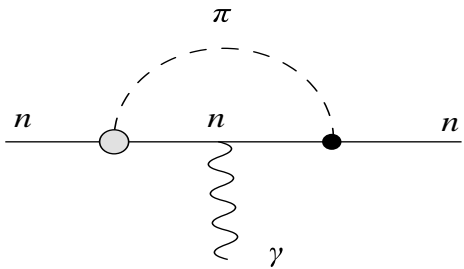}
\end{center}
\caption{A dominant contribution to the neutron EDM through chiral pion exchange. The
shaded blob is CP violating coupling $\bar g_{\pi nn}$, and the black dot is the strong
coupling $g_{\pi nn}$. The neutron couples to photon via its anomalous magnetic moment.}
\label{edm2}
\end{figure}

The neutron EDM from the Feynman diagram in Fig. \ref{edm2} is
\begin{equation}
d_n = \frac{e~g_{\pi n n}~\bar g_{\pi n n}}{8 \pi^2} \frac{\mu_N}{2 m_N} F\left( \frac{m_\pi^2}{m_N^2} \right) \ ,
\end{equation}
where $\mu_N=-1.91$ is the neutron anomalous magnetic dipole moment and the loop function
is
\begin{equation} F(s) = \frac{3}{2} - s - \frac{3s-s^2}{2}\ln s +
\frac{s(5s-s^2)-4s}{2\sqrt{s-s^2/4}} \arctan \frac{\sqrt{s-s^2/4}}{s/2} \ .
\end{equation}
The contribution is suppressed by the mixing angle $\zeta$ between $W_L$ and $W_R$, but
is enhanced by the chiral logarithmic factor~$\displaystyle
\ln\left(\frac{\Lambda_{\chi}}{m_{\pi}}\right)^2$. Putting in all the known physical
parameters, we arrive at an approximate formula
\begin{equation} |d_n^e| \simeq 3\times 10^{-19}
\sin \zeta ~{\rm Im}\left( e^{- i \alpha} V^L_{ud} V_{ud}^{R*} \right)~e{\rm cm} \ ,
\label{nedm}
\end{equation}
which is approximately a function of $r \sin \alpha$ for small $\alpha$.

\begin{figure}[hbt]
\begin{center}
\includegraphics[width=8cm]{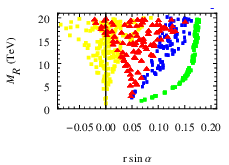}
\includegraphics[width=8cm]{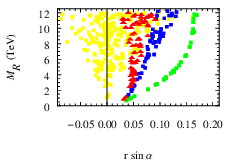}
\end{center}
\caption{Constraints on the mass of $W_R$ and the spontaneous CP violating parameter
$\alpha$ from kaon decay parameter $\epsilon$ ($M_H=\infty$, red triangle; $M_H=75$ TeV,
blue square; $M_H=20$ TeV, large green dots) and neutron EDM (yellow dots). In the right
panel, the theoretical EDM result is reduced by a factor of 5.} \label{edm-epsilon}
\end{figure}

For a fixed $M_H$, the neutron EDM and $\epsilon$ can be used to give a lower bound for
$M_{W_R}$ as well as a corresponding solution for $r \sin\alpha$. On the left panel in
Fig. \ref{edm-epsilon}, we have shown the neutron EDM constraint as a function of
$r\sin\alpha$ and $M_{W_R}$, and it is obvious that the EDM limit prefers small $r \sin
\alpha$. The smallest spontaneous CP phase is obtained when $M_H$ is large and decouples,
and then
\begin{equation}\label{edme}
    M_{W_R} > 8 ~{\rm TeV} \ ,
\end{equation}
which is a very tight bound. At this point, we also fix the product
\begin{equation}\label{ralpha}
   r \sin \alpha \simeq 0.05 \ .
\end{equation}
For a lower $M_{H}$, the spontaneous CP phase must be large from the $\epsilon$
constraint, and the corresponding lower bound on $M_{W_R}$ increases considerably. For
example, when $M_H=75$ TeV, $r\sin\alpha$ is now greater than $0.1$, and the lower bound on
$M_{W_R}$ becomes $\sim $ 18 TeV.

We note that there is considerable hadronic uncertainty in the evaluation of CP-violating
coupling $\bar g_{\pi nn}$ and, to the less extent, chiral perturbation expansion.
However, even one allows a factor of 5 over-estimate in the hadronic calculation, the
combined $\epsilon$ and EDM will still provide a strong constraint on $M_{W_R}$ on the
order of order 4 TeV, as shown on the right panel in Fig. \ref{edm-epsilon}. If $M_H=25$
TeV, the lower bound on $M_{W_R}$ becomes 8 TeV. A future improvement on the neutron EDM
data and theoretical calculation can strengthen this bound considerably.

\begin{figure}[hbt]
\begin{center}
\includegraphics[width=10cm]{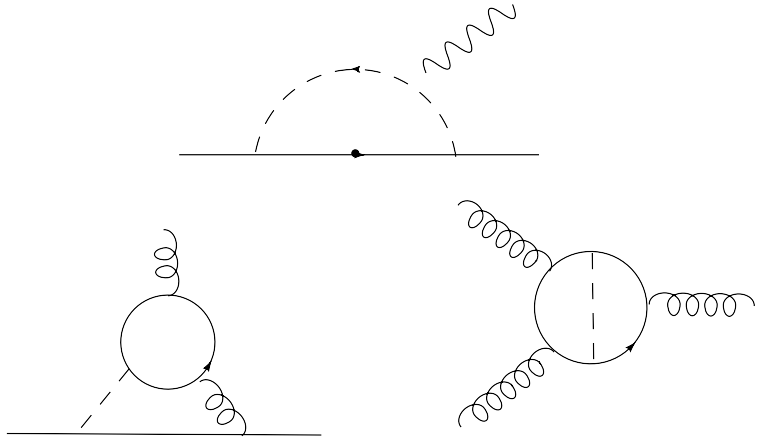}
\end{center}
\caption{Higgs exchange contributions relevant to neutron EDM. The dashed lines include
both charged and neutral Higgs bosons exchanges. The two-loop diagrams contain closed
top-quark loops.} \label{he1}
\end{figure}

We finally comment on the Higgs boson exchange contributions to the neutron EDM.
According to Sec. \ref{coup}, the Higgs bosons $H_1^0$, $A_1^0$ and $H_2^+$ have CP
violating couplings (Eq. (\ref{LN}) and Eq. (\ref{LC})) to the quark fields. The valence
quark EDM receives contributions from virtue $H_1^0$, $A_1^0$ and $H_2^+$ exchange as
shown in Fig.~\ref{he1}. Potentially large contribution also comes from neutral Higgs
boson exchange and virtual top-quark effect at two loops \cite{zee}, as well as two loop
pure gluonic operators due to charged/neutral Higgs exchange \cite{Weinberg}. A complete
analysis of these contributions has been carried out in a pseudomanifest LRS limit with
two doublets Higgs fields instead of triplets considered here \cite{Chang:1992bg}. With
the explicit form of right-handed CKM mixing, we re-evaluate these contributions in the
general case of CP violation. As discussed in Sec. \ref{treele}, $H_1^0$ and $A_1^0$ must
be heavy enough to suppress their contribution to kaon mixing, and we take their masses
to be 15 TeV. In this case, the charged-Higgs $H_2^+$ exchange dominates, whose
contribution to d-quark EDM is approximated by
\begin{equation}
d_d \simeq \frac{2e}{3} \frac{m_t G_F}{4 \sqrt{2} \pi^2} 2 \xi \frac{m_t^2}{m_{H_2^+}^2}
\ln \left( \frac{m_{H_2^+}^2}{m_t^2} \right) \eta_d {\rm Im} \left( V^L_{td} V_{td}^{R*}
e^{-i \alpha} \right) \ ,
\end{equation}
where the scaling factor $\eta_d \simeq 0.12$. The d-quark CDM $f_d$ come from a similar
diagram with photon replaced by gluon leg, $e f_d \simeq
\displaystyle{\frac{3}{2}\frac{\eta_{f}}{\eta_d}} d_d$. The contribution to u-quark EDM
and CDM is suppressed by a factor $\displaystyle \frac{m_b}{m_t}$ and is negligible
compared to that of $d$-quark. Meanwhile, the two loop diagrams are found to be negligibly small.
To a certain level of approximation, the neutron EDM can
be related to the quark EDM and CDM through the $SU(6)$ relation:
\begin{equation}\label{edme}
d_n^e = \frac{1}{3} \left( 4 d_d - d_u \right) + \frac{1}{3} \left( \frac{4}{3} f_d + \frac{2}{3} f_u \right) \ .
\end{equation}
A more accurate relation would use the tensor charges of the nucleon.
\begin{figure}[hbt]
\begin{center}
\includegraphics[width=6cm]{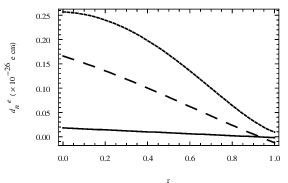}
\includegraphics[width=6cm]{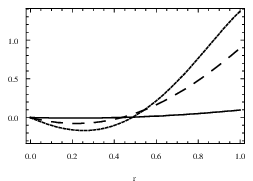}
\end{center}
\caption{One scenario (all $s_q = 1$) for the higgs exchange contribution to neutron EDM
as a function of $r$ for different values $\alpha=0.05$ (solid line), $\alpha=0.5$ (long
dashed line) and $\alpha = 1$ (short dashed line). We choose the FCNH mass to be $15$ TeV
and the charged Higgs mass equal to $3$ TeV.} \label{he2}
\end{figure}

In Fig.~\ref{he2}, we plot the Higgs exchange contributions to the neutron EDM as a
function of $r$ for different values of $\alpha$. We choose the FCNH mass to be $15$ TeV,
and the charged Higgs mass be $3$ TeV. We find the contributions are always smaller than
$10^{-26}e$ cm, well within the experimental bound. Therefore one can neglect the Higgs
exchange contribution without altering the $W$-mass bound for the neutron EDM.

\subsection{Direct CP Violation $\epsilon'$}

The direct CP violation in neutral kaon to $\pi\pi$ decay is calculated via
\begin{equation}
\epsilon'=\frac{i}{\sqrt{2}} \omega \left(\frac{q}{p}\right) \left(\frac{{~\rm{Im}}A_2}
{{~\rm{Re}}A_2}-\frac{{~\rm{Im}}A_0}{{~\rm{Re}}A_0}\right) e^{i(\delta_2-\delta_0)} \ ,
\end{equation}
where the decay amplitudes $A_0$ and $A_2$  are defined as the
matrix elements of the $\Delta S = 1$ effective Hamiltonian between
the neutral-K meson and the isospin $I= 0$ and $2$ $\pi\pi$ states,
\begin{equation}
\langle(2\pi)_I|(-i)H_{\Delta S=1}|K^0\rangle=A_Ie^{i\delta_I}  \ .
\end{equation}
$\delta_I$ is the strong phase for $\pi\pi$ scattering, $\omega \equiv A_2 / A_0$ and
$p$, $q$ are the mixing parameters for $K^0-\overline{K}^0$. To an excellent
approximation, $\omega$ can be taken as real and $q/p = 1$. Therefore, we focus on
calculating the imaginary part of the decay amplitudes.

In the SM, the contributions to $\epsilon'$ come from both QCD and
electromagnetic penguin diagrams \cite{Shifman:1976ge}. The QCD
penguin contributes exclusively to $\Delta I =1/2$ decay, whereas
the electromagnetic penguin is mainly responsible for $\Delta I
=3/2$ decay. Both contributions are important but have opposite
signs. Therefore, the final result depends on precision calculation
of hadronic matrix elements. The state-of-art chiral perturbation
theory \cite{Buras1,Buchalla:1989we,prime,Bosch:1999wr} and lattice
QCD calculations \cite{Blum:2001xb,Pekurovsky:1998jd} have not yet
been sufficiently accurate to reproduce the experimental result
\cite{expprime}. A review of the standard model calculation can be
found in Ref. \cite{Bertolini:2000dy,Buras:2003zz}.

\begin{figure}[hbt]
\begin{center}
\includegraphics[width=10cm]{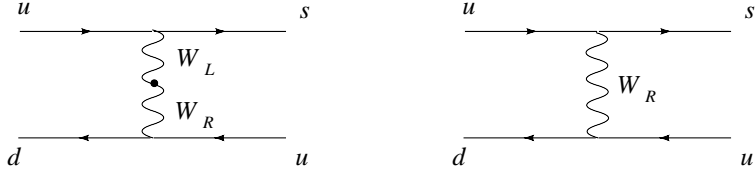}
\end{center}
\caption{New tree-level contribution to the $\Delta S = 1$ interaction from LRSM.}
\label{e'}
\end{figure}

In LRSM, each element in the righthanded CKM matrix has a substantial CP phase. As a
consequence, there are tree level contributions to the phases of $A_2$ and $A_0$.
Following closely the work by Ecker and Grimus \cite{Ecker:1985vv}, the $\Delta S = 1$
effective Hamiltonian from Eq.~(\ref{WW}) and the tree-level Feynman diagram in Fig.
\ref{e'} is
\begin{widetext}
\begin{eqnarray}\label{H}
\mathscr{H}^{\rm tree}_{\Delta S =
1}&=&\sqrt{2}G_F\lambda^{LL}_u\left[\left(\frac{\alpha_S(\mu^2)}
{\alpha_S(M^2_L)}\right)^{-\frac{2}{b}}O^{LL}_+(\mu) +\left(\frac{\alpha_S(\mu^2)}
{\alpha_S(M^2_L)}\right)^{\frac{4}{b}}O^{LL}_-(\mu)\right]\nonumber\\
&+& \sqrt{2}G_F\frac{M^2_L}{M^2_R}\lambda^{RR}_u\left[\left(\frac{\alpha_S(\mu^2)}
{\alpha_S(M^2_R)}\right)^{-\frac{2}{b}}O^{RR}_+(\mu) +\left(\frac{\alpha_S(\mu^2)}
{\alpha_S(M^2_R)}\right)^{\frac{4}{b}}O^{RR}_-(\mu)\right]\nonumber\\
&+&2\sqrt{2}G_F\sin\zeta\lambda^{LR}_ue^{i\alpha}\left[\left(\frac{\alpha_S(\mu^2)}
{\alpha_S(M^2_L)}\right)^{\frac{8}{b}}O^{LR}_-(\mu) -\left(\frac{\alpha_S(\mu^2)}
{\alpha_S(M^2_L)}\right)^{-\frac{1}{b}}O^{LR}_+(\mu)\right]\nonumber\\
&+&2\sqrt{2}G_F\sin\zeta\lambda^{RL}_ue^{- i\alpha}\left[\left(\frac{\alpha_S(\mu^2)}
{\alpha_S(M^2_L)}\right)^{\frac{8}{b}}O^{RL}_-(\mu) -\left(\frac{\alpha_S(\mu^2)}
{\alpha_S(M^2_L)}\right)^{-\frac{1}{b}}O^{RL}_+(\mu)\right] \ ,
\end{eqnarray}
\end{widetext}
where we have taken into account the leading-logarithm QCD corrections with
renormalization scale $\mu$ taken to be around 1 GeV, and $b = 11-2N_f/3$ with $N_f$ the
number of active fermion flavors. The mixing coupling $\lambda^{AB}_u=V^{\rm
CKM*}_{Aud}V^{\rm CKM}_{Bus}$, $A$, $B$ are $L$, $R$. The four quark operators are
\begin{eqnarray}\label{O} O^{LL,RR}_{\pm}&=&\overline{d}\gamma^\mu
\mathbb{P}_{L, R} u\overline{u} \gamma_\mu \mathbb{P}_{L,R}s \pm\overline{d}\gamma^\mu
\mathbb{P}_{L, R}  s \overline{u} \gamma_\mu \mathbb{P}_{L, R}  u \ , \nonumber\\
O^{LR,RL}_+&=&\overline{d}\gamma^\mu  \mathbb{P}_{L, R}
u\overline{u}\gamma_\mu \mathbb{P}_{R,L} s+\frac{2}{3}
\overline{d} \mathbb{P}_{R, L} s\overline{u} \mathbb{P}_{L, R} u \ , \nonumber\\
O^{LR,RL}_-&=&\frac{2}{3}\overline{d} \mathbb{P}_{R, L} s\overline{u} \mathbb{P}_{L, R} u
\ ,
\end{eqnarray}
where $\mathbb{P}_{L, R}$ are projection operators. There are also new penguin diagrams
involving the right-handed gauge boson contributing to $\epsilon'$. However, these
contributions are suppressed by loop factors and are neglected here.

The hadronic matrix elements of the four-quark operators are calculated using the
factorization assumption,
\begin{eqnarray}
\langle(2\pi)_{I=0}| O^{LL,RR}_{+} |\overline K^0\rangle &=& \pm \frac{X}{3 \sqrt{6}} \ , \nonumber \\
\langle(2\pi)_{I=2}| O^{LL,RR}_{+} |\overline K^0\rangle &=& \pm \frac{2 X}{3 \sqrt{3}} \ , \nonumber \\
\langle(2\pi)_{I=0}| O^{LL,RR}_{-} |\overline K^0\rangle &=& \pm \frac{X}{2 \sqrt{6}} \ , \nonumber \\
\langle(2\pi)_{I=2}| O^{LL,RR}_{-} |\overline K^0\rangle &=& 0 \ , \nonumber \\
\langle(2\pi)_{I=0}| O^{LR,RL}_{+} |\overline K^0\rangle &=& \pm \frac{4 X}{9 \sqrt{6}}\ , \nonumber \\
\langle(2\pi)_{I=2}| O^{LR,RL}_{+} |\overline K^0\rangle &=& \pm \frac{2 X}{9 \sqrt{3}}\ , \nonumber \\
\langle(2\pi)_{I=0}| O^{LR,RL}_{-} |\overline K^0\rangle &=& \mp \frac{1}{\sqrt{6}}
\left( \frac{X}{18} + \frac{Y}{2} + \frac{Z}{6} \right) \ , \nonumber \\
\langle(2\pi)_{I=2}| O^{LR,RL}_{-} |\overline K^0\rangle &=& \mp \frac{1}{6\sqrt{3}}
\left( \frac{X}{6} - Z \right) \ ,
\end{eqnarray}
where the parameters $X$, $Y$ and $Z$ are
\begin{eqnarray}
X &\equiv& - \langle \pi^-| \bar d \gamma_\mu \gamma_5 u | 0 \rangle \langle \pi^+ |
\bar u \gamma^\mu s | \overline K^0 \rangle \ , \nonumber \\
&=& i \sqrt{2} F_\pi (m_K^2 - m_\pi^2) \simeq 0.03i~ {\rm GeV}^3 \nonumber \\
Y &\equiv& - \langle \pi^+ \pi^-| \bar u u | 0 \rangle \langle 0 | \bar d \gamma_5 s |
\overline K^0 \rangle \ , \nonumber \\
&=& i \sqrt{2} F_K A^2 (1 - m_K^2 / m_\sigma^2)^{-2} \simeq 0.273i ~{\rm GeV}^3 \nonumber \\
Z &\equiv& - \langle \pi^-| \bar d \gamma_5 u | 0 \rangle \langle \pi^+ | \bar u s |
\overline
K^0 \rangle \nonumber \\
&=& i \sqrt{2} F_\pi A^2 (1 - m_\pi^2 / m_\sigma^2)^{-2} \simeq
0.18i~{\rm GeV}^3\ ,
\end{eqnarray}
where $A = m_K^2 / (m_s + m_d)$, $F_\pi = 93$ MeV and $F_K = 1.22~F_\pi$.

The numerical estimates are taken from Ref. \cite{frere}. $Y$ and $Z$ are much bigger
than $X$ due to chiral enhancement. Clearly the factorization approximation must be
improved as indicated by the empirical $\Delta I=1/2$ rule, which is beyond the scope of
this paper. We note, however, that for our estimation of the bound on $M_{W_R}$, a
multiplicative uncertainty factor on the matrix elements is reduced by a square-root.

\begin{figure}[hbt]
\begin{center}
\includegraphics[width=8cm]{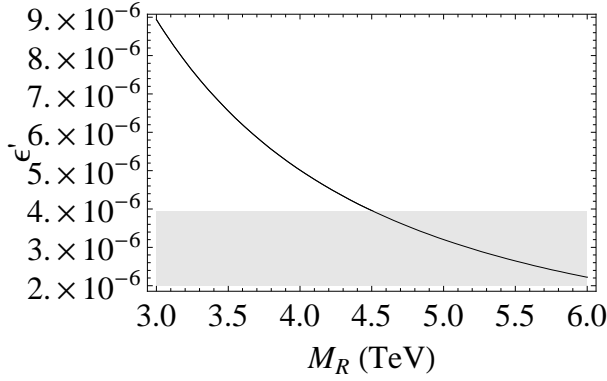}
\end{center}
\caption{$\epsilon'$ as a function of $M_{W_R}$ for $\alpha =
\arcsin \displaystyle \frac{1}{10}$, $r=0.5$ and all $s_q =1$. The shaded part is
allowed by the experimental data}\label{e'2}
\end{figure}

To calculate the weak phases of the decay amplitudes $A_0$ and $A_2$, we use the
experimental value for the real parts of $A_0$ and $A_2$: Re$ A_0 \simeq 3.33\times
10^{-7}$ and $\omega \simeq 1/22$. The dominant new contribution is from the $W_L-W_R$
mixing term due to enhanced hadronic matrix elements and larger CP violation phase
$\alpha$. In fact, because the phase $\alpha$ in the apparent factor $e^{i\alpha}$ is
much larger than the phase in $\lambda_u^{RL,LR}$ which is typically of order $\theta_1$
and $\theta_2$ given in Eq.~(\ref{theta}), $\epsilon'$ is approximately a function of $r
\sin \alpha$, rather than $r$ and $\sin \alpha$ independently. Since $r \sin \alpha$ has
been fixed by $\epsilon$ and $d_n^e$ in the previous subsections, $\epsilon'$ is
approximately a function of $M_{W_R}$ only. In Fig. \ref{e'2}, we plot $\epsilon'$ as a
function of $M_{W_R}$ for $\alpha = \arcsin\displaystyle\frac{1}{10}$, $r=0.5$ and
$s_ds_s =1$ which is required by the neutron EDM calculation. [All $s_i = 1$.] Requiring
that the new contribution should be no larger than $|\epsilon'_{{\rm expt}}| = 3.92
\times 10^{-6}$ \cite{PDG}, we get a lower bound
\begin{equation}
    M_{W_R} > 4.2 ~{\rm TeV} \ ,
\end{equation}
Here we obtain a slightly tighter bound on $M_{W_R}$ than that from kaon mixing. However,
because of the $\Delta I=1/2$ rule, the factorization assumption might have overestimated
the phase of $A_2$. If we take $r\sin\alpha=0.15$, as required by low $M_H$, the bound
changes to 7.4 TeV.

Finally, there are also tree-level FCNH contributions to $H_{\Delta S=1}$, as one can see
from the lagrangian in Eq. (\ref{LN}) \cite{Ecker:1985vv, frere, hehe}. Since the relevant
coupling is suppressed by either the Cabibbo angle or the quark masses, their
contribution is negligible.

\subsection{CP Violation in $B_d\rightarrow J/\psi K_S$ Decay: $S_{J/\psi K}$}

\begin{figure}[hbt]
\begin{center}
\includegraphics[width=10cm]{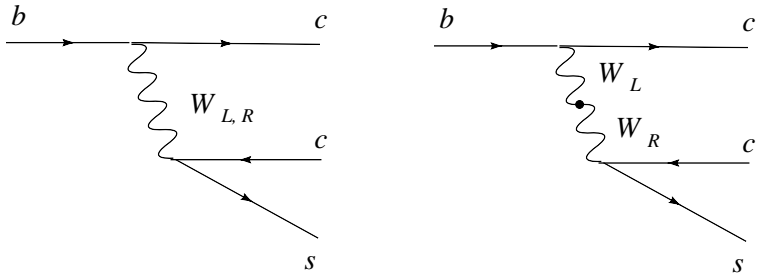}
\end{center}
\caption{Tree level Feynman diagrams for $B_d\rightarrow J/\psi K_s$ from SM and
LRSM.}\label{Jpsi}
\end{figure}

The CP violation in B-meson decay was first observed in $B_d\rightarrow J/\psi K_S$. In
SM, the decay proceeds mainly through the tree-level $b\rightarrow c\bar c s$ and the
penguin contributions are expected to be suppressed by CKM and/or loop factors. The
tree-level diagram is shown in the left panel in Fig.\ref{Jpsi} with an intermediate
$W_L$ exchange.

The relevant CP asymmetry is defined as
\begin{eqnarray}
S_{J/\psi K_S} = \frac{2 ~{\rm Im} \lambda_d}{1 + |\lambda_d|^2} \ ,
\end{eqnarray}
where
\begin{eqnarray}
\lambda_{d} = \left(\frac{q}{p}\right)_B \cdot \frac{\mathscr{A}(\bar{B_d} \rightarrow
J/\psi K_S)} {\mathscr{A}(B_d \rightarrow  J/\psi K_S)} \ , \nonumber
\end{eqnarray}
$\mathscr{A}$ is a decay amplitude and $(q/p)_B$ is from the B-meson mixing.

The magnitude of $\lambda_d$ is close to 1 and thus $S_{J/\psi K_S}\sim {\rm
Im}\lambda_d$. In the SM, $S_{J/\psi K_S}$ predominantly comes from $B-\bar B$ mixing,
and is related to the $\beta$ angle of the unitary triangle because the ratio of the
decay amplitude is independent of the hadronic matrix element when the tree operator
dominates. Experimentally, $\sin 2 \beta_{{\rm expt}} = 0.673 \pm 0.028$ \cite{PDG}.

In the LRSM, the effective $\beta$ angle will receive new contributions from initial and
final neutral meson mixings \cite{scpv} and from the new tree-level decay operators
through $W_R$ exchange and $W_R-W_L$ mixing, as shown in Fig. \ref{Jpsi}. We will not
consider the kaon mixing contribution for the following reason: Since $K_S$ is dominantly
CP-even $|K_S\rangle =p|K^0\rangle + q|\overline K^0\rangle$ and $\overline B^0$ decay
involves $\overline K^0$, $\lambda_d$ is proportional to $(q/p)_K^*$. The imaginary part
of $(q/p)_K^*$ is proportional to the imaginary part of $\epsilon$ which is known to be
on the order of $10^{-3}$, much smaller than the phase in $(q/p)_B$. The $\epsilon$
constraint on the LR symmetric model has already been studied independently and we
decouple the kaon-mixing effect from $S_{J/\psi K_S}$. Thus we write
\begin{eqnarray}
2 \beta^{eff} \approx && 2 \beta  +  {~\rm arg} \left( 1 + \frac{M_{12}^{B_d,
LR}}{M_{12}^{B_d, SM}} \right) \nonumber \\
&& + {~\rm arg} \left( 1 + \frac{\langle J/\psi \overline K_0|\mathscr{H}^{LR}|\overline
B_0\rangle }{\langle J/\psi \overline K_0|\mathscr{H}^{SM}|\overline B_0\rangle}
\right)\left( 1 + \frac{\langle J/\psi K_0|\mathscr{H}^{LR}| B_0\rangle }{\langle J/\psi
K_0|\mathscr{H}^{SM}| B_0\rangle} \right)^{-1} \ .
\end{eqnarray}
The $\mathscr{H}^{LR}$ operator is similar to that in Eq.~(\ref{H}), with substitutions
$s\rightarrow b$, $u\rightarrow c$, and $d\rightarrow s$.

The ratio ${M_{12}^{B_d, LR}}/{M_{12}^{B_d, SM}}$ can be calculated from Eq.~(58) and a
corresponding expression from the SM. Its magnitude is around $10^{2}M_{W_L}^2/M_{W_R}^2$
and carries a phase factor $V_{td}^{R*}V_{tb}^R/V_{td}^{L*}V_{tb}^L= s_ds_b
e^{-i(\theta_1+\theta_3)}$. The hadronic matrix element for the decay is less known. In
the naive factorization approximation for $\langle J/\psi \overline
K_0|\mathscr{H}^{SM}|\overline B_0\rangle$, the decay rate is under predicted by an order
of magnitude \cite{Cheng}. Therefore, the non-factorization contribution must be
significant \cite{li}. In the ratio of matrix elements, we expect the factorization
approach work better. Using this approximation, we find
\begin{equation}
\frac{\langle J/\psi \overline K_0|\mathscr{H}^{LR}|\overline B_0\rangle }{\langle J/\psi
\overline K_0|\mathscr{H}^{SM}|\overline B_0\rangle} = \frac{M_{W_L}^2}{M_{W_R}^2}~
\frac{\lambda_c^{RR}}{\lambda_{c}^{LL}}~ \frac{2 \left[ \alpha_S(m_b)/ \alpha_S(M_{W_L})
\right]^{-2/b} - \left[ \alpha_S(m_b)/ \alpha_S(M_{W_L}) \right]^{4/b}}{2 \left[
\alpha_S(m_b)/ \alpha_S(M_{W_R}) \right]^{-2/b} - \left[ \alpha_S(m_b)/ \alpha_S(M_{W_R})
\right]^{4/b}} \ ,
\end{equation}
where $\lambda_c^{AB} = V^{A*}_{cs} V^B_{cb}$ for $A, B = L, R$, so
$\lambda_c^{RR}/\lambda_c^{LL} = s_s s_b e^{-i (\theta_2 + \theta_3)}$. The
non-perturbative $B\rightarrow K$ form factors $F_+(m_{J/\psi}^2)$ has been canceled out
and the result is independent of hadronic parameters. The magnitude of the ratio is
${\cal O}(M_{W_L}^2/M_{W_R}^2)$ and hence is much smaller than the mixing contribution to
$2\beta_{\rm eff}$. This conclusion remains valid even if we underestimate this ratio by
an order of magnitude.

\begin{figure}[hbt]
\begin{center}
\includegraphics[width=8cm]{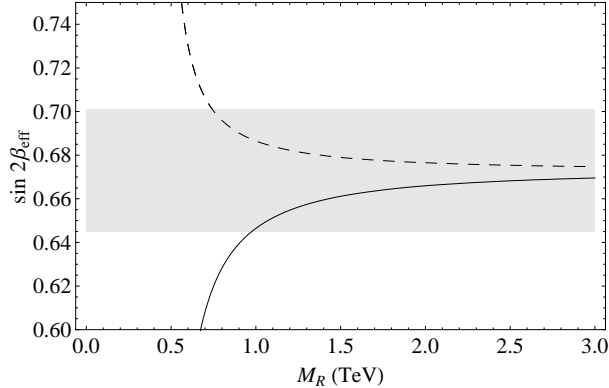}
\end{center}
\caption{Predicted CP asummetry in $B^0_d \rightarrow J/\psi K_S$ as a function of
$M_{W_R}$, for $r \sin \alpha = 0.05$. The solid line corresponds to the sign choice
$s_i=1$ and the dashed line corresponds to $s_i=1$ except $s_t=-1$. The shaded part is
allowed by the experimental data. }\label{jpsi}
\end{figure}

The modified CP asymmetry in LRSM depends on the righthanded scale $M_{W_R}$ and
$r\sin\alpha$. We take $r \sin \alpha \simeq 0.05$ as determined from the $\epsilon$
constraint. There are two independent choices for quark mass signs which generate
different predictions. We take either $s_t=+1$ or $s_t=-1$, and the results are shown in
Fig. \ref{jpsi}. Demanding $\sin 2 \beta_{eff}$ to lie within the experimental error bar,
we get a moderate lower bound on $M_{W_R}$,
\begin{equation}
    M_{W_R} > 0.7 ~{\rm TeV} \ .
\end{equation}
As we have commented above, this constraint comes predominantly from the $B_d-\overline
B_d$ mixing contribution.

\section{Conclusion}

In this paper, we have made a comprehensive study of CP violating observables in the
low-energy sector of the minimal LRSM with the only assumption of parity invariance
imposed on the theory. This is made possible by an explicit solution for the right-handed
quark mixing matrix with explicit dependence on the spontaneous CP violation phase
$\alpha$. Although the hadronic physics uncertainty is still large, the CP observables do
provide significant and strong constraint on the right-handed $W$-boson mass scale
$W_{M_R}$ and the FCNH mass scale $M_H$. In fact, a new experiment result and/or improved
theoretical calculation on EDM might provide the strongest bound yet on the right-handed
gauge boson mass.

\begin{table}[hbt]
\begin{tabular}{|c||c|c|c|}
\hline
&~$M_{W_R}$~(TeV)~&~$M_{H_1^0},~M_{A_1^0}$~(TeV)~&~$~\left|r\sin\alpha\right|~$  \\
\hline \hline
$\Delta m_K$~&~2.5~&~15~&~$-$  \\
\hline
$\Delta m_{B_d}$~&~2.5~&~12~&~$-$ \\
\hline
$\Delta m_{B_s}$~&~2.7~&~25~&~$-$ \\
\hline &~4~(8)~&~$^*$100~&~0.05 \\ \cline{2-4}
~\raisebox{2.3ex}[0pt]{$d_n^e$~\&~$\epsilon$}~~&~8~(20)~&~$^*$25~&~0.15 \\
\hline & 4.2 & $-$  & $^*$0.05 \\ \cline{2-4}
\raisebox{2.3ex}[0pt]{$\epsilon'$} & 7.4 & $-$ & $^*$0.15 \\
\hline & 0.8 & $-$  & $^*$0.05 \\ \cline{2-4}
\raisebox{2.3ex}[0pt]{$B \rightarrow J_\psi K_S$} & 1.3 & $-$ & $^*$0.15 \\
\hline

\end{tabular}
\caption[]{A summary of bounds on $M_{W_R}$ and $ M_H$ from different physical
observables. Stars on the items indicate input values. For the neutron EDM case, the
numbers in the parenthesis are direct result from Eq. (\ref{nedm}) and those without are
obtained by reducing theoretical values by a factor of 5.}\label{table2} \label{table1}
\end{table}

We stress the point that CP must be violated both explicitly and spontaneously in the
minimal model, with one bidoublet and two triplets. Up to ${\cal O}(\lambda^3)$, we can
write $V_R^{CKM}$ in a compact form as in Eqs. (\ref{vr}), (\ref{vr1}) and (\ref{theta}).
We find that $V_R$ has the same hierarchical structure as $V_L$, i.e. elements of the two
mixing matrices are suppressed by the same orders of Cabibbo angle $\lambda = 0.22$. And
because of the spontaneous CP phase $\alpha$, each element in $V_R^{CKM}$ acquires a
phase angle proportional to $r \sin \alpha$. Therefore, the phenomenology related to CP
violation in the minimal LRSM turns out to be very rich. We explored mass differences of
neutral kaon and B-meson with updated lattice results, and found an updated lower bound
on righthanded $W$-boson mass: $M_{W_R}>2.5$ TeV. With the CP violating processes, we
find a combined bound $M_{W_R}> 4\sim 8$ TeV from $\epsilon$ and neutron EDM constraints
when the FCNH contribution is ignored. And we can also fix $r \sin \alpha \simeq 0.05$
from the combined bound. When the FCNH contribution is added to $\epsilon$, the bound is
stronger when the $M_H$ becomes small. We go on to study $\epsilon'$ and CP asymmetry in
$B_d\rightarrow J/\psi K_S$ decay. By applying the experimental constraints and
theoretical uncertainty, we conclude that the lower bound on righthanded $W$-boson mass
surviving from all above experimental constraint is about 4 TeV. If the $W_R$ really have
a mass close to this lower bound, it is possible to detect its signal in the up coming
LHC.

We also find that the lower bound on $M_{H_1^0}>25$ TeV is tighter than the bound
previous bounds \cite{bounds, Pospelov:1996fq}. Perhaps, this suggests that one should
have two bidoublets so that one can invoke cancelation between them, as in the
spontaneous CP violation model discussed in Ref. \cite{ylwu}. In that case for our
analytic solution to remain valid, the second bi-doublet should develop vev. All bounds
are shown in Table \ref{table2} for easy reading of the results of the paper.

Finally, we would like to comment on the constraint on $r$ arising from the consideration
of the mass shift of SM Higgs boson in LRSM. According to discussions in Sec.
\ref{higgs3}, the SM Higgs mass is
\begin{equation}
m^2_{h}=\left(4\lambda_1-\frac{\alpha^2_1}{\rho_1}\right)\kappa^2+\alpha_3\xi^2v_R^2 \ ,
\end{equation}
to second order in $\alpha$, $\xi$ and $\epsilon$.  The shift in mass due to LRS is
$\displaystyle \alpha_3\xi^2v_R^2 - \frac{\alpha^2_1}{\rho_1} \kappa^2$ and can be
expressed in terms of the masses of FCNH
\begin{equation}\label{51}
\Delta m_{h^0} \simeq \xi M_{H_1^0} \ .
\end{equation}
From the discussions below Eq.~(\ref{28}), FCNH masses have to be large enough to
suppress the tree level contribution to kaon mixing. On the other hand, the SM Higgs mass
should not exceed TeV scale in order to preserve perturbative unitarity. A recent
analysis~\cite{Djouadi:2005gi} yields
\begin{equation}
m^2_{h^0} < 870{~\rm GeV} \ .
\end{equation}
The lower bound on the FCNH boson mass $M_{H_1^0}>25$ TeV yields an upper bound on $r=\xi
m_t/m_b<1.44$. From discussions on CP violating observables, $0.05<|r\sin\alpha|<0.15$,
this translates into a lower bound on the spontaneous phase $ |\alpha| > 0.035 $.

This work was partially supported by the U. S. Department of Energy via grant
DE-FG02-93ER-40762. Y. Z. acknowledges the hospitality and support from the TQHN group at
University of Maryland and a partial support from NSFC grants 10421503 and 10625521. X.
J. is supported partially by a ChangJiang Scholarship at Peking University. R. N. M. is
supported by NSF grant No. PHY-0652363.


\begin{thebibliography}
\frenchspacing
\bibitem{lrmodel}
R.~N.~Mohapatra and J.~C.~Pati,
  Phys.\ Rev.\  D {\bf 11}, 566 (1975);
R.~N.~Mohapatra and J.~C.~Pati,
  Phys.\ Rev.\  D {\bf 11}, 2558 (1975);
G.~Senjanovic and R.~N.~Mohapatra,
  Phys.\ Rev.\  D {\bf 12}, 1502 (1975); Phys. Rev.\ D {\bf 23}, 165
(1981);
For a review, Rabindra N. Mohapatra, {\it CP Violation}, World Scientific Publ. Co., C.
Jarlskog, Ed., 1989.

\bibitem{lee}
T. D. Lee, talk given at the Center for High-Energy Physics, Peking
University, Nov. 2006.

\bibitem{scpv} R.~N.~Mohapatra, F.~E.~Paige and D.~P.~Sidhu,
  Phys.\ Rev.\  D {\bf 17}, 2462 (1978);
D.~Chang,
  Nucl.\ Phys.\  B {\bf 214}, 435 (1983);
G.~C.~Branco, J.~M.~Frere and J.~M.~Gerard,
  Nucl.\ Phys.\  B {\bf 221}, 317 (1983);
G.~C.~Branco and L.~Lavoura,
  Phys.\ Lett.\  B {\bf 165}, 327 (1985).

\bibitem{generalcp}
P.~Langacker and S.~Uma Sankar,
  Phys.\ Rev.\  D {\bf 40}, 1569 (1989);
G.~Barenboim, J.~Bernabeu, J.~Prades and M.~Raidal,
  Phys.\ Rev.\  D {\bf 55}, 4213 (1997).

\bibitem{kiers}
K.~Kiers, J.~Kolb, J.~Lee, A.~Soni and G.~H.~Wu,
  Phys.\ Rev.\  D {\bf 66}, 095002 (2002).

\bibitem{gorbahn} G.~Barenboim, M.~Gorbahn, U.~Nierste and M.~Raidal,
  Phys.\ Rev.\  D {\bf 65}, 095003 (2002).

\bibitem{ourpaper}
  Y.~Zhang, H.~An, X.~Ji and R.~N.~Mohapatra,
  Phys.\ Rev.\  D {\bf 76}, 091301 (2007).

\bibitem{marshak} R.~E.~Marshak and R.~N.~Mohapatra, Phys.\
Lett.\  B {\bf 91}, 222 (1980).

\bibitem{PDG}
  W.~M.~Yao {\it et al.}  [Particle Data Group],
  J.\ Phys.\ G {\bf 33}, 1 (2006).

\bibitem{pot}
N.~G.~Deshpande, J.~F.~Gunion, B.~Kayser and F.~I.~Olness,
  Phys.\ Rev.\  D {\bf 44}, 837 (1991).

\bibitem{pot1}
K.~Kiers, M.~Assis and A.~A.~Petrov,
  Phys.\ Rev.\  D {\bf 71}, 115015 (2005).

\bibitem{masiero} A.~Masiero, R.~N.~Mohapatra and R.~D.~Peccei,
  Nucl.\ Phys.\  B {\bf 192}, 66 (1981).

\bibitem{wolf} J.~Basecq, J.~Liu, J.~Milutinovic and L.~Wolfenstein,
  Nucl.\ Phys.\  B {\bf 272}, 145 (1986).

\bibitem{pseudo} G.~Ecker and W.~Grimus,
  Phys.\ Lett.\  B {\bf 153}, 279 (1985);
G.~Barenboim, J.~Bernabeu and M.~Raidal,
  Nucl.\ Phys.\  B {\bf 478}, 527 (1996);
P.~Ball, J.~M.~Frere and J.~Matias,
  Nucl.\ Phys.\  B {\bf 572}, 3 (2000).

\bibitem{frere}
  J.~M.~Frere, J.~Galand, A.~Le Yaouanc, L.~Oliver, O.~Pene and J.~C.~Raynal,
  Phys.\ Rev.\  D {\bf 46}, 337 (1992).

\bibitem{Duka:1999uc}
  P.~Duka, J.~Gluza and M.~Zralek,
  Annals Phys.\  {\bf 280}, 336 (2000)
  [arXiv:hep-ph/9910279].

\bibitem{soni}
G.~Beall, M.~Bander and A.~Soni,
  Phys.\ Rev.\ Lett.\  {\bf 48}, 848 (1982).

\bibitem{neubert}
  M.~Neubert,
  Z.\ Phys.\  C {\bf 50}, 243 (1991).

\bibitem{bounds}
R.~N.~Mohapatra, G.~Senjanovic and M.~D.~Tran,
  Phys.\ Rev.\  D {\bf 28}, 546 (1983);
 G.~Ecker, W.~Grimus and H.~Neufeld,
  Phys.\ Lett.\  B {\bf 127}, 365 (1983)
  [Erratum-ibid.\  B {\bf 132}, 467 (1983)];
F.~J.~Gilman and M.~H.~Reno,
  Phys.\ Rev.\  D {\bf 29}, 937 (1984);
S.~Sahoo, L.~Maharana, A.~Roul and S.~Acharya,
  Int.\ J.\ Mod.\ Phys.\  A {\bf 20}, 2625 (2005).




\bibitem{kaon2}
D.~Chang, J.~Basecq, L.~F.~Li and P.~B.~Pal,
  Phys.\ Rev.\  D {\bf 30}, 1601 (1984);
W.~S.~Hou and A.~Soni,
  Phys.\ Rev.\  D {\bf 32}, 163 (1985);
J.~Basecq, L.~F.~Li and P.~B.~Pal,
  Phys.\ Rev.\  D {\bf 32}, 175 (1985).


\bibitem{Bfactor}
R.~Babich, N.~Garron, C.~Hoelbling, J.~Howard, L.~Lellouch and
C.~Rebbi,
  Phys.\ Rev.\  D {\bf 74}, 073009 (2006);
D.~J.~Antonio {\it et al.},
  arXiv:hep-lat/0702026.




\bibitem{running}
A.~J.~Buras, S.~Jager and J.~Urban,
  Nucl.\ Phys.\  B {\bf 605}, 600 (2001).


\bibitem{Ecker:1985vv}
  G.~Ecker and W.~Grimus,
  Nucl.\ Phys.\  B {\bf 258}, 328 (1985).

\bibitem{Buras1}
  A.~J.~Buras,
  arXiv:hep-ph/9806471.

\bibitem{Djouadi:2005gi}
  A.~Djouadi,
  arXiv:hep-ph/0503172.



\bibitem{lattice}
  N.~Tantalo,
  arXiv:hep-ph/0703241.

\bibitem{lattice2}
  D.~Becirevic {\it et al.},
  Nucl.\ Phys.\  B {\bf 634}, 105 (2002)
  [arXiv:hep-ph/0112303].


D.~Becirevic, V.~Gimenez, G.~Martinelli, M.~Papinutto and J.~Reyes,
  JHEP {\bf 0204}, 025 (2002)
  [arXiv:hep-lat/0110091];
H.~Wittig,
  Eur.\ Phys.\ J.\  C {\bf 33}, S890 (2004)
  [arXiv:hep-ph/0310329].


\bibitem{Pospelov:1996fq}
  M.~E.~Pospelov,
  Phys.\ Rev.\  D {\bf 56}, 259 (1997)
  [arXiv:hep-ph/9611422].

\bibitem{edm}
G.~Beall and A.~Soni,
  Phys.\ Rev.\ Lett.\  {\bf 47}, 552 (1981);
 G.~Ecker, W.~Grimus and H.~Neufeld,
  Nucl.\ Phys.\  B {\bf 229}, 421 (1983);
J.~M.~Frere, J.~Galand, A.~Le Yaouanc, L.~Oliver, O.~Pene and
J.~C.~Raynal,
  Phys.\ Rev.\  D {\bf 45}, 259 (1992).

\bibitem{hehe}
X.~G.~He, B.~H.~J.~McKellar and S.~Pakvasa,
  Phys.\ Rev.\ Lett.\  {\bf 61}, 1267 (1988).

\bibitem{he}
  X.~G.~He and B.~McKellar,
  Phys.\ Rev.\  D {\bf 47}, 4055 (1993).

\bibitem{dynamics}
J.~Donoghue, E.~Golowich and B.~Holstein, 
{\it Dynamics of the Standard Model}~(Cambridge University Press, Cambridge, 1994).

\bibitem{zee}
S.~M.~Barr and A.~Zee,
  Phys.\ Rev.\ Lett.\  {\bf 65}, 21 (1990)
  [Erratum-ibid.\  {\bf 65}, 2920 (1990)].


\bibitem{Weinberg}
S.~Weinberg,
  Phys.\ Rev.\ Lett.\  {\bf 63}, 2333 (1989).


\bibitem{Chang:1992bg}
  D.~Chang, X.~G.~He, W.~Y.~Keung, B.~H.~J.~McKellar and D.~Wyler,
  Phys.\ Rev.\  D {\bf 46}, 3876 (1992)
  [arXiv:hep-ph/9209284].


\bibitem{Shifman:1976ge}
  M.~A.~Shifman, A.~I.~Vainshtein and V.~I.~Zakharov,
  Sov.\ Phys.\ JETP {\bf 45}, 670 (1977)
  [Zh.\ Eksp.\ Teor.\ Fiz.\  {\bf 72}, 1275 (1977)].


\bibitem{Buchalla:1989we}
  G.~Buchalla, A.~J.~Buras and M.~K.~Harlander,
  Nucl.\ Phys.\  B {\bf 337}, 313 (1990).



\bibitem{prime}
G.~Buchalla, A.~J.~Buras and M.~E.~Lautenbacher,
  Rev.\ Mod.\ Phys.\  {\bf 68}, 1125 (1996)
  [arXiv:hep-ph/9512380].

\bibitem{Bosch:1999wr}
  S.~Bosch, A.~J.~Buras, M.~Gorbahn, S.~Jager, M.~Jamin, M.~E.~Lautenbacher and L.~Silvestrini,
  Nucl.\ Phys.\  B {\bf 565}, 3 (2000)
  [arXiv:hep-ph/9904408].

\bibitem{Blum:2001xb}
  T.~Blum {\it et al.}  [RBC Collaboration],
  Phys.\ Rev.\  D {\bf 68}, 114506 (2003)
  [arXiv:hep-lat/0110075].

\bibitem{Pekurovsky:1998jd}
  D.~Pekurovsky and G.~Kilcup,
  Phys.\ Rev.\  D {\bf 64}, 074502 (2001)
  [arXiv:hep-lat/9812019].

\bibitem{expprime}
  H.~Burkhardt {\it et al.}  [NA31 Collaboration],
  Phys.\ Lett.\  B {\bf 206}, 169 (1988);
  V.~Fanti {\it et al.}  [NA48 Collaboration],
  Phys.\ Lett.\  B {\bf 465}, 335 (1999)
  [arXiv:hep-ex/9909022];
  A.~Alavi-Harati {\it et al.}  [KTeV Collaboration],
  Phys.\ Rev.\ Lett.\  {\bf 83}, 22 (1999)
  [arXiv:hep-ex/9905060].

\bibitem{Bertolini:2000dy}
  S.~Bertolini, J.~O.~Eeg and M.~Fabbrichesi,
  Phys.\ Rev.\  D {\bf 63}, 056009 (2001)
  [arXiv:hep-ph/0002234].

\bibitem{Buras:2003zz}
  A.~J.~Buras and M.~Jamin,
  JHEP {\bf 0401}, 048 (2004)
  [arXiv:hep-ph/0306217].










\bibitem{Cheng}
J.~Chay and C.~Kim, [arXiv:hep-ph/0009244];
H.~Cheng and K.~Yang, Phys.\ Rev.\ D {\bf 63}, 074011 (2001)
[arXiv:hep-ph/0011179].

\bibitem{li}
C.~Chen and H.~Li,
Phys.\ Rev.\ D {\bf 71}, 114008 (2005)
[arXiv:hep-ph/0504020].






\bibitem{ylwu}
Y.~L.~Wu and Y.~F.~Zhou, arXiv:0709.0042 [hep-ph].


\end{thebibliography}
\end{document}